\documentclass[aps,prd,amssymb,eqsecnum,nofootinbib,superscriptaddress]{revtex4}
\usepackage{epsfig}
\usepackage{amssymb}
\usepackage{amsmath}
\usepackage{amsfonts}
\usepackage{bm}
\newcommand{\be}{\begin{equation}}
\newcommand{\ee}{\end{equation}}
\newcommand{\bea}{\begin{eqnarray}}
\newcommand{\eea}{\end{eqnarray}}
\newcommand{\bes}{\begin{subequations}}
\newcommand{\ees}{\end{subequations}}
\newcommand{\nn}{\nonumber}

\begin{document}

\title{Gaussianity of  LISA's confusion backgrounds}
\author{\'{E}tienne Racine}
\affiliation{Department of Physics, Mathematics and Astronomy, California Institute of Technology, Pasadena, CA 91125}
\author{Curt Cutler}
\affiliation{Jet Propulsion Laboratory, California Institute of Technology, Pasadena, CA 91109}
\date{today}
\begin{abstract}
Data analysis for the proposed Laser Interferometer Space Antenna (LISA) will be complicated by the 
huge number of sources in the LISA band.   In the frequency band $\sim 10^{-4}-2\times 10^{-3}\,$Hz, galactic white dwarf  binaries (GWDBs) are sufficiently dense in frequency space that it will be impossible to resolve most of them, and "confusion noise" from the unresolved 
Galactic  binaries will dominate over instrumental noise in determining LISA's sensitivity to other sources in that band.  Confusion noise from unresolved extreme-mass-ratio inspirals (EMRIs) could also contribute significantly to LISA's  total noise curve.  To date, estimates of the effect of LISA's confusion noise on matched-filter searches and their detection thresholds have generally approximated the noise as  Gaussian,  based on the Central Limit Theorem. However in matched-filter searches,  the appropriate detection threshold for a given class of signals may be located rather far out on the tail of the signal-to-noise probability distribution, where a priori it is unclear whether the Gaussian approximation is reliable.
Using the Edgeworth expansion and the theory of large deviations, we investigate the probability distribution of the usual matched-filter detection statistic,  far out on the tail of the distribution.  We apply these tools to four somewhat idealized versions of LISA data searches: searches for
EMRI signals buried in GWDB confusion noise, 
and searches for massive black hole binary (MBHB) signals buried in i) GWDB noise, ii) EMRI noise, and iii) a sum of EMRI noise and Gaussian noise.
Assuming reasonable short-distance cut-offs in the populations of confusion sources (since the
very closest and hence strongest sources will be individually resolvable),
modifications to the appropriate detection threshold, due to the non-Gaussianity of the confusion noise, turn out to be quite small for realistic cases.
The smallness of the correction is partly due to the fact that these three types of sources  evolve on quite different timescales, so no single background source closely resembles any search template. We also briefly discuss other types of LISA searches where the non-Gaussianity of LISA's confusion backgrounds could perhaps have a much greater impact on search reliability and efficacy.
\end{abstract}
\pacs{04.25.Nx,04.30.Db,04.80.Nn,95.75.Wx,95.85.Sz} %taken from CC's BBO paper
\maketitle

\section{Introduction}
Data analysis for the proposed Laser Interferometer Space Antenna (LISA) will be complicated by the 
huge number of sources in the LISA band.  For example, while LISA is expected to detect 
of order $10^4$ individual compact binaries (especially white dwarf-white dwarf binaries) in our Galaxy,  in the frequency band $\sim 10^{-4}-2\times 10^{-3}\,$Hz such binaries are sufficiently dense in
frequency space that it will  be impossible to resolve most of them. The 
"confusion noise" from all the unresolved Galactic  binaries will dominate over
instrumental noise in determining LISA's sensitivity to other sources in that band.
Extreme-mass-ratio inspirals (EMRIs) are another very important category of LISA sources. 
EMRIs are  inspirals of stellar-mass compact objects (white dwarfs, neutron stars, or black holes) into
massive ($\sim 10^6 M_{\odot}$) black holes (MBHs) in galactic nuclei.  Because of their extremely
small mass ratio, EMRI sources remain in the LISA band for timescales of order years.  
While LISA can do a great deal of interesting science with individually detected EMRIs, the 
EMRIs that are too faint to be resolved also constitute a confusion background, partially masking other
sources.  Barack and Cutler~\cite{BC2} (hereinafter BC2) estimated the spectral density of confusion noise from unresolved EMRIs and found that it becomes comparable to that of LISA's instrumental noise  or WD confusion noise
only if EMRI event rates turn out to be at the high end of the estimated range.  BC2 therefore concluded that LISA's EMRI confusion 
background would be rather benign:
either the EMRI rates are low-to-medium, in which case non-EMRI noise
sources dominate the total noise, or the EMRI rates are high, in which case noise from unresolvable EMRIs could dominate, 
but the EMRI detection rate is {\it also} higher (which would more than compensate, from a scientific standpoint).

However there is a potential caveat to BC2's treatment of EMRI  confusion noise (as well as to many discussions of the white dwarf
confusion noise) related to the Gaussianity of that noise.
Because the number of undetected GWDBs  or EMRIs will be large ($\sim 10^7-10^8$ for GWDBs and $\sim 10^5 -10^6$ for EMRIs), the confusion background has generally been treated as approximately Gaussian, based on an appeal to the Central Limit Theorem. However in matched-filter searches,  the detection threshold for a given class of signals may be located rather far out on the tail of the signal-to-noise probability distribution. For example,  in searching for EMRIs,  the vast number of independent EMRI signals that can be searched for necessitates a detection threshold of $\sim 14\,\sigma$~\cite{Gairetal}, assuming Gaussian statistics;  similarly we estimate
that searches for MBHBs will require a signal-to-noise detection threshold of $\sim 7 \,\sigma$, to keep false alarms at an acceptable level.
Then naturally one must confront the question:
How much does the tail of the distribution of the usual detection statistic
deviate from Gaussian in the range $\sim 7 -14 \sigma$? In other words, how often does the confusion background manage to mimic the signals searched for, at the level of the usual detection threshold?  Or put yet another way: how much {\it higher} must one set the detection threshold
to compensate for the non-Gaussianity of the tail of the distribution?
We analyse these questions by harnessing two tools from statistics, namely the
Edgeworth expansion and the theory of large deviations, and applying them to model problems that are somewhat idealized versions of the cases that will arise in actual LISA data analysis.

In this paper we will be concerned with three types of LISA sources, all of which are binaries: GWDBs, MBHBs, and EMRIs.
To make the calculations below analytically tractable, we shall assume that the binary orbits are quasi-circular (i.e., circular except for a slow inspiral due to gravitational radiation reaction), and we shall approximate each gravitational waveform by its lowest-order piece in a post-Newtonian expansion. Also, while LISA should return two independent science data channels (and a third at high frequency), for simplicity
we shall treat the output as a single channel.
 Other simplifications and approximations are discussed below.

To avoid confusion, we should emphasize that in this paper we are mainly interested in detections near the threshold SNR. Now, the strongest MBHB signals detected by LISA (perhaps from $\sim 10^6 M_{\odot}$ MBHs merging at redshift $z<1$) will likely have (matched-filtering) SNRs $\sim 10^3$; however
those are {\it not} the MBHB signals that interest us here.  Instead, when we discuss searches for
MBHB signals, we are mainly interested in the
most distant resolvable ones; e.g., mergers of $\sim 10^4 M_{\odot}$ MBHs at $z \sim 20$.
Likewise when we discuss searches for  EMRI signals, our interest is in the weakest resolvable
ones.   Of course, the overall detection rates will likely be dominated by the weakest detectable
signals.

This paper is structured as follows. We first discuss, in section \ref{sec:foundations}, the role of confusion noise in a matched filter search and show how it reduces to a statistical problem involving sums of independent identically distributed random variables. We then describe in some detail the two statistical techniques we apply in this paper, namely the Edgeworth expansion and the theory of large deviations. Next, in section \ref{sec:chirps}, we describe in detail our toy models for the GWDB and EMRI confusion noise respectively. In particular we discuss our model waveforms, which are simply Newtonian circular-orbit chirps, and our choices for binary parameter distributions. Finally in section \ref{sec:apps} we use
confusion noise models of section \ref{sec:chirps} 
and apply the tools described in section \ref{sec:foundations} 
to four model searches: searches for EMRIs signals imbedded in 
GWDB confusion noise and searches for MBHBs imbedded in i) GWDB noise, ii) EMRI noise, and
iii) the sum of EMRI noise  and Gaussian noise (instrumental plus GWDB), respectively.
In each case we obtain the probability distributions for the usual detection statistic and
assess the impact of the non-Gaussianity of the confusion noise on the appropriate detection
threshold. Our conclusions are summarized in section V.  In an appendix we present  a heuristic derivation of 
the central result in large-deviations theory  (Chernoff's formula), describe its relation to the Edgeworth expansion, and apply it to a 
a simple, illustrative case--the binomial distribution.

Throughout this paper, we use geometrical units in which $G=c=1$. Therefore everything can be measured in our fundamental unit of seconds.   For familiarity, we sometimes express 
quantities in terms of yr, Mpc, or $M_\odot$, which are related to our fundamental unit by 1 yr $= 3.1556 \times 10^7$ s, 1 Mpc $= 1.029 \times 10^{14}$ s, and $1 M_\odot = 4.926 \times 10^{-6}$ s.

\section{Statistical foundations}\label{sec:foundations}

\subsection{Confusion noise in matched-filter searches}

As an introduction to the general problem of searching for gravitational-wave (GW) signals that may be buried in confusion noise, 
consider a LISA data set $s(t)$ that 
%consider a matched-filter search for some signal $h(t)$ in a LISA data set $s(t)$ 
is dominated by instrumental noise plus unresolved background signals, but which may also contain some resolvable signal proportional to $h(t)$; i.e., 
\be\label{noise}
s(t) = n(t) + \sum_{i=1}^N h_i(t) + \rho \, h(t)
%s(t) = n(t) + \sum_{i=1}^N h_i^{bg}(t) + \rho h(t)
\ee
where $n(t)$ represents (Gaussian) instrumental noise, the sum over $N$ sources represents the confusion background, $h(t)$ is the sought-for signal [normalized to $(h \big| h) = 1$, where the inner product $( \ \  \big| \ \ ) $ is defined below], 
and $\rho$ represents the overall strength of the sought-for signal.  If $h(t)$ is simply not present in the data, then $\rho = 0$.
For example, the background signals $h_i(t)$ could be from GWDBs~\footnote{Since the WD binaries are located much closer to us than the sought-for MBHB, their summed signal is sometimes called  "foreground" instead of "background". However we shall refer to all confusion noise populations simply as "backgrounds".}, while $h(t)$ is  the gravitational wave signal from some MBHB.
In a matched-filter search for $h(t)$ in this data set, one basically just computes the inner product $(s|h)$:

\be
(s|h) = (n|h) + \sum_i^N x_i + \rho \, ,
%(s|h) = (n|h) + X
\ee
%\be
%X \euqiv  \sum_i^N x_i \, .
%\ee
where $(n|h)$ is a Gaussian random variable, and where
each $x_i  \equiv (h_i|h)$ is a random variable drawn from some probability distribution function (PDF) $p(x)$. 
There must be {\it some} threshold value $\rho_{th}$, such that when $(s|h) > \rho_{th}$ one can claim a detection with
very high confidence (say, $> 99\%$).  But what is this threshold value?  To compute $\rho_{th}$, we
need to know the probability distribution of the sum
\be\label{defX}
X \equiv  \sum_i^N x_i  \, .
\ee
Most of the work in this paper will be spent in estimating the probability distribution function for $X$, 
$P_N(X)$, given its parent distribution $p(x)$. 
 We will be particulary concerned with the behavior of 
$P_N(X)$ at large $X$ -- out on the "high-$\sigma$" tail.
In the next subsection we describe two tools from statistics that are quite useful in this context.      

\subsection{The Central Limit Theorem and Beyond}
Let $p(x)$ be some normalized PDF.
The $q^{\rm th}$ moment of $p(x)$ is defined to be
\be\label{moment}
\mu_q \equiv E[x^q] = \int{ x^q p(x) dx}.
\ee
%For convenience (and without loss of generality) we shall assume that $\mu_1$ (the mean value of $x$) vanishes.
We shall assume for convenience that $\mu_1$ (the mean value of $x$) vanishes, since it automatically does
so in all applications in this paper.
We shall also assume that the second moment $\mu_2$ (the variance of $x$) exists, and define $\sigma_x \equiv \sqrt{\mu_2}$. 
%Assuming the second moment $\kappa_2$ (the variance of $x$) also exists, 
Let $Z$ be the average value of  $N$ samples from this distribution:
\be\label{Zdef}
Z \equiv \frac{1}{N}\sum_{i=1}^N x_i
\ee
Then the Central Limit Theorem basically states that 
in the limit of large $N$, the PDF for $Z$, $P_N(Z)$, approaches a Gaussian with variance $\mu_2/N$.
Defining the re-scaled variable $Y \equiv \frac{\sqrt N}{\sigma_x} Z$, we have 
%standard deviations $\sigma_x/\sqrt{N}$:
%\be\label{PNZ}
%P_N(Z) \rightarrow \left(\frac{N}{2\pi \sigma_x^2}\right)^{1/2}\exp\left[-\frac{N Z^2}{2\sigma_x^2}\right]
%\ee
\be\label{PNY}
P_N(Y) \rightarrow  \frac{1}{\sqrt{2\pi}} \exp\left[-\frac{ Y^2}{2}\right]  \, 
\ee
%where $\sigma_x^2 \equiv \kappa_2$.
as $N \rightarrow \infty$.

While the Central Limit Theorem states that  $P_N(Y)$ converges to a Gaussian for large $N$, for this paper it is crucial to realize that the convergence of the {\it ratio} $P_N(Y)/[(2\pi)^{-1/2} e^{-Y^2/2}]$ to unity can be remarkably slow at large values of $Y$.
This is particularly true if some higher moments of $p(x)$ diverge, as happens, e.g., if $p(x)$ has only power-law decrease at large $x$.

To quote standard theorems on the convergence of $P_N(Y)$ to a Gaussian, we need a few more definitions.
Define $F_N(Y)$ to be the cumulative distribution function (CDF) of $P_N(Y)$, 
\be\label{cdf}
F_N(Y) = \int_{-\infty}^{Y} P_N(\tilde Y)\, d\tilde Y\, ,
\ee
and let $\bar F_N(Y)$ be the complementary function to $F_N(Y)$:
\be
\bar F_N(Y) = \int_{Y}^{\infty} P_N(\tilde Y)\, d\tilde Y\,  = 1 - F_N(Y) \, .
\ee
Also define $\Phi(Y)$ to be the CDF of a Gaussian,
\be
\Phi(Y) \equiv \frac{1}{\sqrt{2\pi}}\int_{-\infty}^{Y} e^{-\tilde Y^2/2} d\tilde Y \, .
\ee
and let 
\be
\bar \Phi(Y) \equiv \frac{1}{\sqrt{2\pi}}\int_{-\infty}^{Y} e^{-\tilde Y^2/2} d\tilde Y \,  = 1 - \Phi(Y) \, .
\ee
Of course, $\bar\Phi(Y) = (1/2)\, {\rm erfc}(Y/\sqrt{2})$, where "erfc" is the complementary error function. 

Now let us further assume that the absolute third moment $\rho_3 \equiv E[|x|^3] > 0$ of the parent distribution $p(x)$ exists and is finite.
For this case, a well-known result on the convergence of (\ref{PNY})
is the Berry-Ess\'een Theorem, which states that for all $Y$ and $N$, 
\be\label{Berry}
\sup_Y \bigg| F_N(Y)  - \Phi(Y)\bigg| \le C \frac{\rho_3}{\sigma_x^3} N^{-1/2} \, .
\ee
where $C$ is some constant less than 0.7655~\cite{Feller,Shiganov}. 
Of course, this is equivalent to 
\be\label{cBerry}
\sup_Y \bigg|  \bar F_N(Y)  - \bar \Phi(Y)\bigg| \le C \frac{\rho_3}{\sigma_x^3} N^{-1/2} \, .
\ee

Now let us consider the practical implications of the Berry-Ess\'een Theorem. 
%Consider the case $N = 10^4$, and ask 
What threshold value $Y_{th}$ ensures that, say, $\bar F_N(Y_{th}) < 10^{-6}\,$?
Since $\bar\Phi(4.8916) = 10^{-6}$, a first estimate based on the Central Limit Theorem would be
 $Y_{th} \approx 4.8916$. However, by Eq.(\ref{cBerry}), the {\it error}
in this estimate (for $\rho_3/\sigma_x^3$ of order one) can be of order $N^{-1/2}$. 
So the potential error in the Gaussian
estimate greatly exceeds that estimate itself unless $N > 10^{12}$!   
More generally,  for large $Y$, $N$ must be exponentially large --
of order $e^{Y^2}$ --for the right-hand side of (\ref{cBerry}) to be smaller than $\bar\Phi(Y)$.

When higher moments of $p(x)$ exist, one can systematically improve on the Central Limit Theorem estimate of $P_N(Y)$. 
These improvements are described in the next two subsections.

\subsection{The Edgeworth expansion}\label{subsec:edgeworth}

The key ingredient in constructing the Edgeworth expansion is the cumulant generating functional of a PDF, defined as
\be\label{cgen}
\lambda(\omega) \equiv \ln E[e^{i\omega x}] = \ln \int_{-\infty}^\infty e^{i\omega x} p(x) \, dx.
\ee
%Since all raw moments of $p(x)$ are well-defined, 
One can expand the exponential and then the logarithm about $\omega = 0$ in (\ref{cgen}) to obtain the following series for the cumulant generating functional:
\be
\lambda(\omega) = \sum_{q=2}^\infty \frac{\kappa_q}{q!}(i\omega)^q,
\ee
where $\kappa_q$ is called the $q^{\rm th}$ cumulant of the parent distribution. Now consider the cumulant generating functional $\Lambda(\omega)$ of $P_N(Y)$, which is given by
\bea
\Lambda(\omega) &=&  \ln E[e^{i\omega Y}] \nn \\
&=& N \lambda\left(\frac{\omega}{\sqrt{N\sigma_x^2}}\right) \nn \\
&=& N \sum_{q=2}^\infty \frac{\kappa_q/\sigma_x^q}{q!}\left(\frac{i\omega}{\sqrt{N}}\right)^q. \label{cYgen}
\eea
where $\sigma_x^2 = \kappa_2$.  Notice that 
%the cumulant generating functional for $P_N(Y)$
$\Lambda(\omega)$ 
depends only on $N$ and the cumulants of $p(x)$. Taking the exponential of both sides of (\ref{cYgen}), formally expanding the results around $\omega = 0$, and then gathering terms according to powers of $N^{-1/2}$ yields

\be\label{PYft}
E[e^{i\omega Y}] = e^{-\omega^2/2}\left[1 + \sum_{r=1}^\infty \frac{P_r(i\omega)}{N^{r/2}}\right],
\ee
where $P_r(i\omega)$ is a polynomial in $i\omega$ depending only on the cumulants $\kappa_q$. Since the left-hand side of (\ref{PYft}) is simply the Fourier transform of $P_N(Y)$, taking the inverse transform on both sides of (\ref{PYft}) finally gives

\be\label{Edgeworth}
P_N(Y) = \frac{1}{\sqrt{2\pi}}e^{-Y^2/2}\left[1 + \sum_{r=1}^\infty \frac{Q_r(Y)}{N^{r/2}}\right],
\ee
where the $Q_r(Y)$ are polynomials in $Y$. The first few terms of the series are 

\be\label{Edgeworthexplicit}
P_N(Y) = \frac{1}{\sqrt{2\pi}}e^{-Y^2/2}\left[1 + \frac{\kappa_3\, / \sigma_x^3}{6\sqrt{N}}H_3(Y) + \frac{\kappa_4\, / \sigma_x^4}{24N}H_4(Y) + \frac{\kappa_3^2 / \sigma_x^6}{72N}H_6(Y) + O(N^{-3/2})\right],
\ee
where $H_q(Y)$ is the Chebyshev-Hermite polynomial of order $q$, the ones appearing above being %equal to
\bes
\bea
H_3(Y) &=& Y^3 - 3Y \\
H_4(Y) &=& Y^4 - 6Y^2 +3 \\
H_6(Y) &=& Y^6 - 15Y^4 + 45Y^2 - 15 \, .
\eea
\ees

At this point we should emphasize that while the Edgeworth series (\ref{Edgeworth}) is formally correct, it does not converge in general. Rather, in the limit $N \rightarrow \infty$ it provides an asymptotic expansion of the true CDF  $F_N(Y)$.  More precisely, let  $\Lambda_r(Y)$ be related to the polynomials $Q_r(Y)$ defined above by
\be
\Lambda_r(Y) = \int_{-\infty}^{Y}\frac{e^{-{\tilde Y}^2/2}}{\sqrt{2\pi}}Q_r({\tilde Y})d{\tilde Y}.
\ee
Assume that the first $k$ cumulants $\kappa_q$ exist, for some $k \ge 3$.   Also assume that $\lim_{T\rightarrow \infty} \sup_{|t| > T} |v(t)|  <1$, where
\be\label{vdef}
v(t) \equiv \int e^{i t x} p(x) dx \, .
\ee
(This condition on $v(t)$ will be easily satisfied for all parent distributions $p(x)$ we consider.)
Then Theorem 3 in section VI of Petrov~\cite{Petrov} states that
\be\label{Petrov3}
\lim_{N\rightarrow \infty} N^{(k-2)/2} \bigg(1 + |Y|^k\bigg) \bigg|F_N(Y)  - \Phi(Y) - \sum_{r = 1}^{k-2} N^{-r/2} \Lambda_r(Y)\bigg| = 0
\ee
uniformly in $Y$ ($-\infty < Y < \infty$).

Assuming that $p(-x) = p(x)$ (as will be true for all examples considered in this paper), so that the
odd cumulants of $p(x)$ all vanish, and assuming the first $k$ cumulants exist (for $k$ even), then this theorem implies that 
the error in the $(k-2)^{\rm th}$-order approximation to $F_N(Y)$ scales like 
$N^{-k/2}$ as $N \rightarrow \infty$. 
For example, assuming $\kappa_4$ %\sigma_x^4$ 
exists, the error in the second-order Edgeworth expansion of 
$F_N(Y)$ scales like $N^{-2}$ for sufficiently large $N$.
E.g., assuming $\kappa_4/\sigma_x^4$ is of order $1$, 
one therefore generally requires $N > 10^{8}$ for this potential error to be smaller than $10^{-16}$.
(Again, if one uses  $\sim 10^{14}$ independent templates in the search , then one would want the false alarm probability for any one of them to be smaller than $\sim 10^{-16}$.)

Now fix $Y$ and $N$.  Since the Edgeworth series is only asymptotic, one will typically find that the first few terms in the series might get smaller and smaller, and their sum ever closer
to $P_N(Y)$, but eventually the terms in the series may start to grow and the sum diverges. 
A useful rule of thumb is then to truncate the Edgeworth expansion before the first term that is larger than the previous ones.

If not all moments $\kappa_q$ exist, it becomes clear why the  Edgeworth series cannot converge, since all terms  in the expansion decrease exponentially with $Y$ at large $Y$, while 
$F_N(Y)$ falls off much more slowly. To see this, consider the case where $p(x)$ is an even function 
 having a power-law tail:
\be\label{powerlaw}
p(x) \rightarrow B \sigma_x^m x^{-m-1}\ \ {\rm for}\,\, |x| \gg \sigma_x    %\ \ \  {\rm at\,\, large}\,\, x
\ee
for some constant $B$ and some odd $m>0$. Let $f(x)$ be the CDF for $p(x)$, and let $\bar f(x) \equiv 1- f(x)$.
Then clearly $\bar f(x) \rightarrow (B/m)  (x/\sigma_x)^{-m}$ at large $x$.  Now fix the number of samples, $N$.  Following
Bazant~\cite{Bazant}, we note that the probability that 
the sum 
$\sum_{i=1}^N (N^{1/2} \sigma_x)^{-1} x_i$ is greater than some value $Y$ is
clearly of the same order or greater than the probability that any single term in the sum is greater than $Y$,
so
\be 
\bar F_N(Y) \gtrsim N \bar f(N^{1/2} \sigma_x Y) \rightarrow (B/m)  N^{1- \frac{m}{2}}Y^{-m}
\ee
at large $Y$.   So if the parent distribution has a power-law tail, then for any fixed $N$, $P_N(Y)$ has the same 
power-law fall-off at very large $Y$.  (Of course, the above argument just shows that $\bar F_N(Y)$  falls  off no 
faster than $Y^{-m}$; however it seems likely that $\bar F_N(Y)$ and $\bar f(x)$ fall off according to the same
power law at large $Y$ and $x$, respectively~\cite{Bazant}.)

This line of reasoning  suggests that the Edgeworth expansion becomes unreliable
at values of $Y$ such that 
\be
(B/m)  N^{1- \frac{m}{2}}Y^{-m} > Y^{-1}e^{-Y^2/2}
\ee
or 
\be\label{Edgest}
Y^2 > (m-2)\,{\rm ln}N + 2(m-1)\,{\rm ln}Y  -2{\rm ln} (B/m) \, .
\ee
We shall typically be interested in cases where $N$ is large enough that the 
$(m-2){\rm ln}N$ dominates the right-hand side of (\ref{Edgest}).
In that case, we obtain the rule of thumb that the Edgworth expansion (and its first term, the Central Limit
Theorem estimate) become unreliable for $Y > \sqrt{(m-2) {\rm ln}N}$.  So the range of 
validity of the Edgeworth expansion increases only like the square root of the exponent $m$ describing
the power law fall-off of $p(x)$.   

The situation changes dramatically if the parent distribution $p(x)$ falls to zero expontially (or faster) 
as $x \rightarrow \infty$.  In that case large-deviations theory guarantees that $P_N(Z)$ also has
exponential fall-off as 
$Z \rightarrow \infty$.  We turn to this subject next.

\subsection{The theory of large deviations}\label{subsec:ld}

The goal of large-deviations theory is to determine the PDF of the random variable $Z$, defined above in Eq.(\ref{Zdef}), on the high-$\sigma$ tails. From the parent distribution $p(x)$, one begins by defining a modified cumulant generating functional $\lambda(\beta)$ as
\be\label{lambda-def}
\lambda(\beta) = \ln \int e^{\beta x} p(x) dx \, .
\ee
[Comparing with (\ref{cgen}), we see that this is simply the usual cumulant generating functional evaluated at imaginary frequency $\omega = -i\beta$.] 
Note that this integral does not exist unless $p(x)$ falls to zero exponentially fast as $x \rightarrow \infty$.  To emphasize this point:  if $p(x)$ has a power-law tail as $x \rightarrow \infty$, then $\lambda(\beta)$ does not exist and the results of large-deviations theory do not apply. In the rest of this subsection, we will assume $p(x)$ is sufficiently well behaved at large $x$ that $\lambda(\beta)$ exists. Then the basic result of large-deviations theory is a theorem due to Cram\'{e}r (e.g., see \cite{Stroock}), which states that 
\be\label{largedev1}
 \lim_{N \rightarrow \infty} {1\over N}\,  {\rm ln}\, \bar F_N(Z) \rightarrow  -I(z) \, ,
\ee
where 
\be\label{IZ}
I(Z) = \max_{\beta}[ Z \beta - \lambda(\beta) ] \, .
\ee
\noindent
This basically implies that for large $N$ and arbitrary $Z$, the PDF of the random variable $Z$ is well approximated by
\be\label{largedev1}
P_N(Z) \approx  C \exp[-N\,  I(Z)]   \, 
\ee
%as $N \rightarrow \infty$, 
where  $C$ is a normalization constant.
%where the function $I(Z)$, called the "rate function", is independent of $N$ at fixed $Z$ and
Eq.~(\ref{IZ}) is sometimes referred to in the literature as Chernoff's formula, and 
the function $I(Z)$ is called the "rate function".  Clearly $I(Z)$ is the Legendre transform of $\lambda(\beta)$. 

There are well-known, close connections between Chernoff's formula and statistical mechanics.
 Roughly, $\beta$ is like an inverse-temperature, $\lambda(\beta)$ is analogous to the Helmholz free energy, and $-N \, I(Z)$ is analogous to the entropy.  
 A gentle introduction to large-deviations theory is given in Ref.\cite{Lewis}. 
 Since we presume most of our readers are unfamiliar with large deviations theory,  we also give a short, heuristic derivation of Eqs.~ (\ref{IZ})  and (\ref{largedev1}) in Appendix \ref{LDapp}.

\section{Confusion noise from populations of binaries}\label{sec:chirps}

In this paper, all the GW sources we consider are types of binaries: WD binaries, MBHBs, and EMRIs.
We shall be considering the problems of searching for one type of binary in the
confusion noise produced by a large number of unresolved sources of a different type; e.g., considering
the search for MBHBs buried in the confusion background of unresolved WDs or unresolved  
EMRIs. Since this paper represents a first-cut at the problem of estimating the non-Gaussian
tails of the detection statistic, we shall simplify the analysis by approximating all three types of 
binaries as  being in (non-precessing) quasi-circular orbits.  We further approximate the emitted gravitational waveform as a simple chirp, with instantaneous frequency $f$ equal to twice the orbital frequency. 
Also, while the waveform that LISA actually measures is modulated (on a 1-year timescale) by the satellite constellation's rotational and translational motion, for simplicity we neglect these modulations.
It should be clear from the derivations, however, that including LISA's orbital modulations would have very little impact on our basic results. 
In the next subsection we briefly describe our model gravitational waveforms and their overlaps.

\subsection{The waveforms from circular-orbit binaries and their overlaps }

Using the quadrupole formula, one can show that (to lowest order in a post-Newtonian expansion)
the instantaneous gravitational-wave frequency  $f$ evolves in time according to
%\be\label{omegaorbdot}
%\dot{\omega}_{\text{orb}} = \frac{96}{5}\mu M^{2/3} \omega_{\text{orb}}^{11/3}.
%\ee
\be\label{omegaorbdot}
\dot f  = \frac{96}{5} \pi^{8/3} \mu M^{2/3} f^{11/3} \, .
\ee
\noindent This  is easily integrated to give

%\be\label{omegaorbt}
%\omega_{\text{orb}}(t) = \omega_{\text{orb}}(0)\left[1 - \frac{t}{t_{\text{rr}}}\right]^{-3/8},
%\ee

\be\label{omegaorbt}
f(t) = f(0)\left[1 - \frac{t}{t_{\text{rr}}}\right]^{-3/8},
\ee
where the radiation reaction timescale $t_{rr}$ is given by

%\be\label{trrdef}
%t_{\text{rr}} = \frac{5}{256}\frac{1}{\omega_{\text{orb}}(0)^{8/3} \mu M^{2/3}}.
%\ee

\be\label{trrdef}
t_{\text{rr}} = \frac{5}{256}\frac{1}{\pi^{8/3} f(0)^{8/3} \mu M^{2/3}}.
\ee

We assume that the gravitational wave strain $h_i(t)$ detected by LISA due to a binary (labeled by $i$) located at distance $D_i$ from the solar system assumes the form 

\bea
h_i(t) &=& A_0 \frac{\mathcal{M}_i}{D_i}\left[\pi \mathcal{M}_i \, f_i(t) \right]^{2/3}\cos\left[\varphi_i + 2\pi\int_0^t f_i(t^\prime)dt^\prime \right] \label{simplifiedstrain}
\eea
where 
$\varphi_i$ is a random initial phase and
$A_0$ is an overall factor of order one (discussed below).  
Again,  $f_i(t) = f_i(0)(1 - t/t_{{\rm rr},i})^{-3/8}$, and henceforth we will refer to $f_i(0)$ as simply $f_i$. 
The quantity $\mathcal{M}_i$ is the binary chirp mass defined as 

\be
\mathcal{M}_i = M_i^{2/5}\mu_i^{3/5},
\ee
where $M_i$ and $\mu_i$ are the (locally measured) total and reduced masses respectively. 

Equation (\ref{simplifiedstrain}) is valid as long as the binary is close enough so that cosmological effects can be neglected. This is certainly true for galactic white dwarf binaries. However this is generally not the case for EMRIs or MBHBs, which will typically be at cosmological distances. For cosmologically distant binaries, we instead have~\cite{markovic}:

\bea
h_i(t) &=& A_0 \frac{\mathcal{M}_i(z_i)}{D_L(z_i)}\left[\pi \mathcal{M}_i(z_i) \, f_i(t) \right]^{2/3}\cos\left[\varphi_i + 2\pi\int_0^t f_i(t^\prime)dt^\prime \right]. \label{simplifiedstrainII}
\eea
\noindent 
The quantity $\mathcal{M}_i(z_i)$ is the redshifted chirp mass defined as 

\be
\mathcal{M}_i(z_i) = (1+z_i) M_i^{2/5}\mu_i^{3/5}.
\ee
The quantity $D_L(z)$ is the standard luminosity distance at redshift $z$ for a flat universe, namely

\be
D_L(z) = \frac{(1+z)}{H_0}\int_0^z \frac{dz^\prime}{[\Omega_m(1+z^\prime)^3 + \Omega_\Lambda]^{1/2}}.
\ee
In this paper we use the values $\Omega_m = 0.3$ and $\Omega_\Lambda = 0.7$. (lt turns out that we do not require a precise value
for $H_0$ for our analyses, since this factor just gets absorbed into a quantity representing the total number of EMRIs out to 
some maximum redshift.)

In reality the overall factor $A_0$ depends on the four angles in the problem (the source's sky location and orientation), and is in fact time-varying due to LISA's changing antenna pattern. 
However  for this paper we neglect those dependencies--in effect approximating $A_0$ by its rms value.  We would not expect this approximation 
to greatly affect the overall shape  of 
$p(x)$.  Moreover, we expect that at large $x$, $p(x)$ is dominated by background sources that are
close (small $D_i$), rather than ones with particularly favorable orientations.  Since it is primarily the tail of $p(x)$
that determines the behavior of $P_N(X)$ at large $X$, we do not expect this averaging over angles to 
greatly affect our conclusions.

We write the confusion noise strain $c(t)$ as follows
\bea
c(t) &=& \sum_{i} h_i(t) \nn \\ 
&=& \sum_{i} A_i(t) \cos\left[\varphi_i + 2\pi\int_0^t f_{i}(t^\prime) dt^\prime \right],
\eea
where
\be\label{Aidef}
A_i(t) = A_0 \frac{\mathcal{M}_i}{D_{i}}\left[\pi \mathcal{M}_i f_i(t)\right]^{2/3}.
\ee
For our purpose, the quantity of interest is the overlap $X \equiv (c | h)$ of a given normalized template $h$ (from a given class of sought-for sources) with the confusion noise $c$:
\be 
(c|h) = \sum_i (h_i|h) \, ,
\ee
where the inner product $(h_i|h)$ is defined as
\be\label{overlapdef}
(h_i|h) =  2 \int_{-\infty}^{+\infty} \frac{\tilde{h_i}(f) \tilde{h}^\ast(f)}{S_n(|f|)}\,df,
\ee
where $\tilde{h_i}(f)$ and $\tilde{h}(f)$ are the Fourier transforms of $h_i(t)$ and $h(t)$, respectively,
and $S_n(|f|)$ is the one-sided noise spectral density. 
In all searches we consider in this paper, the template will also be a Newtonian chirp of the form (\ref{simplifiedstrain}) 
\bea
h(t) &=& A(t) \cos\left[\varphi + 2\pi\int_0^t f(t^\prime) dt^\prime \right] \nn \\
&=& 2\sqrt{\frac{6 \mu M^{2/3}}{5\pi I}}[\pi f(0)(1 - t/t_{\rm rr})^{-3/8}]^{2/3}\cos\left[\varphi + 2\pi\int_0^t f(t^\prime) dt^\prime \right], \label{templatedef}
\eea
where the normalization condition $(h|h) = 1$ implies 
\be\label{Idef}
I = \int_0^\infty \frac{f^{-7/3} \,df}{S_n(f)}.
\ee

Both  $h_i(t)$ and $h(t)$ are instantaneously monochromatic signals with slowly varying frequencies.
Consider the tracks $f_i(t)$ and $f(t)$ that their frequencies sweep out in the $t-f$ plane.
The integral in Eq.~(\ref{overlapdef}) is dominated by the point where the two tracks cross.
Using the stationary phase approximation, the integral can be approximated as~\cite{CurtBBO} 
\be\label{analyticoverlap}
%X = \sum_i \frac{1}{S_n[f_i(t_i)]} A_i(t_i) A(t_i) |\delta \dot{f}_{i}(t_i)|^{-1/2} \cos[\delta \Phi_{i} + \text{sgn}(\delta \dot{f}_{i}) \pi/4],
(h_i|h) = \frac{1}{S_n[f_i(t_i)]} A_i(t_i) A(t_i) |\delta \dot{f}_{i}(t_i)|^{-1/2} \cos[\delta \Phi_{i} + \text{sgn}(\delta \dot{f}_{i}) \pi/4],
\ee
where 

\be
\delta \dot{f}_{i} = \big[\dot{f}(t_i) - \dot{f}_{i}(t_i)\big] = \frac{3}{8}\frac{f(t_i)^{11/3}}{\pi^{8/3}}\left[\frac{1}{t_{\rm rr}f^{8/3}(0)} - \frac{1}{t_{{\rm rr},i}f_i^{8/3}(0)}\right]
\ee
and
\be
\delta \Phi_{i} = \varphi - \varphi_i + 2\pi\int_0^{t_i} \big[f(t^\prime)  - f_{i}(t^\prime)\big] dt^\prime. 
\ee
The time $t_i$ is the instant of time when the template and the $i$th binary cross in the time-frequency plane, i.e. when $f(t_i) = f_{i}(t_i)$. (If the template and the $i^{\rm th}$ binary either do not cross in the time-frequency plane, or cross outside LISA's sensitivity band or outside the observation period, then
we approximate their overlap by zero.  In the applications below this is implemented by restricting the integration range over binary parameters.) 
%Given a set of binary parameters characterizing all the binaries producing the confusion noise, Eq.(\ref{analyticoverlap}) gives a very accurate analytic approximation to the signal-to-noise. 

A GW background is essentially a distribution of unresolved signals.  
In the next two subsections we introduce model distributions for the GWDB and EMRI backgrounds, respectively.

\subsection{Binary parameters for galactic white dwarf binaries}

Here we present our model distribution for the galactic WD binaries.  For simplicity, we
%We next discuss how to generate a given realization of the white dwarf confusion noise.
will assume that all WD binaries have the same chirp mass, for which we adopt the median value arising
from recent population synthesis calculations:
%$M_c = 0.5224 M_\odot$. (This is the $M_c$ for  a binary composed of two $0.6 M_\odot$ WDs.)  
$M_c = 0.25 M_\odot$~\cite{Edlund}. (This is approximately the $M_c$ for  a binary composed of two $0.3 M_\odot$ WDs.)  
We further assume that the other binary parameters are drawn 
from the following distributions:
\bes\label{probs}
\bea
p(D_i)dD_i &=&\theta(D_i-D_{\rm min}(f_i))\, \theta(D_{\rm max}-D_i) \,\frac{2D_i}{D_{\rm max}^{2} - D_{\rm min}^{2}(f_i)} dD_i , \label{Dprob} \\
p(f_i)df_i &=& \theta(f_i - f_{\rm min})\, \theta(f_{\rm max} - f_i)
\, \frac{8}{3f_{\rm min}}\left(\frac{f_i}{f_{\rm min}}\right)^{-11/3}\, df_i, \label{fprob} \\
p(\varphi_i)d\varphi_i &=& \frac{d\varphi_i}{2\pi} , \label{varphiprob}
\eea
\ees
where, in (\ref{fprob}), $f_i$ is the initial (i.e., at the beginning of the data set) gravitational-wave frequency of the binary, 
and $f_{\rm min}$ and  $f_{\rm max}$
%is a low frequency threshold below which no white dwarf binaries can overlap with the template, i.e. 
%essentially the lower bound of the LISA band. 
represent some low- and high-frequency cut-offs for the population we are considering.
For our applications, we shall generally take $f_{\rm min} = 10^{-4}$Hz and $f_{\rm max} = 10^{-2}$Hz. 
The scaling $p(f_i) \propto f_i^{-11/3}$ just comes from the assumption that binaries are
"born" at frequencies below $f_{\rm min}$ and then evolve according to Eq.~(\ref{omegaorbdot}).  
(Basically, binaries  evolve much faster at higher frequency, and so are correspondingly sparser there.)

The distance probability distribution $p(D_i)$ assumes that all galactic WD binaries are uniformly distributed in a disk of radius $D_{\rm max} = 10$kpc, centered on our Solar System. Clearly this inaccurate in two ways. First, the Solar System is not at the center of the Milky Way (nor is the Milky Way a uniform disk).  However the non-uniformity and non-centeredness (around us) of the galactic disk clearly
mostly affects the distribution of binaries more distant than a few $kpc$, and these are
not among the strongest sources. The distant binaries do not strongly affect the "high-$\sigma$" 
tail of $p(x)$, which is what is crucial for determining the tail of $P_N(X)$. Therefore this aspect of the uniform disk assumption should be fairly harmless. More problematic is that disk model
%Clearly this ignores the actual position of the Solar System in the Galaxy, which clearly reduces the 
%number of outliers, i.e. high-$\sigma$ overlaps. 
departs significantly from reality  at distances less than the thickness of the disk, which is $\sim 600$ pc.
Below this distance, it would be better to approximate the WD distribution as spherical.
%the distribution of white dwarf binaries is approximately spherical. 
However we shall not do this for the following reason. Assuming that the distibution is planar
clearly overestimates the number of nearby WD binaries, which artificially amplifies the
high-$\sigma$ tail of $p(x_i)$ (since the closest background sources have the largest
overlaps with any searched-for signal).  At the end of our analysis, we shall find that, even
with the uniform-disk distribution, the non-Gaussianity of the WD background is a negligible factor
in searching for MBHBs or EMRIs.  Had we correctly modified $p(D_i)$ for $D < 600$ pc, the
non-Gaussianity would still be negligible--even more so.  That is, while the uniform-disk assumption
is hard to justify {\it a priori}, it is fully justified {\it a posteriori}.

Finally, we discuss the
inner cutoff $D_{\rm min}(f_i)$.  The justification for imposing an inner cut-off $D_{min}(f_i)$ is that
within this distance any WD binary of frequency $f_i$ would be so bright that it could immediately
be found in the data and essentially subtracted out, {\it before} searching for other types of signals.
%is essentially the maximum distance at which galactic white dwarf binaries can be subtracted out even before the initial MBHB search.
So when we state results, "the GWDB background" is really short for "the GWDB background minus the 
very brightest, immediately identifiable GWDB sources".

What is a reasonable value for $D_{min}(f_i)$?
A straightforward calculation shows that the 
(sky-averaged) LISA signal-to-noise for a white dwarf binary at distance 
$D$ is
\be
%{\rm SNR} = 4.31 \left(\frac{\cal M}{0.5224 M_{\odot}}\right)^{5/3} \left(\frac{1\, {\rm kpc}}{D}\right)
{\rm SNR} = 1.26 \left(\frac{\cal M}{0.25 M_{\odot}}\right)^{5/3} \left(\frac{1\, {\rm kpc}}{D}\right)
\left(\frac{T_{obs}}{3 {\rm yr}}\right)^{1/2}\left(\frac{f}{1 {\rm mHz}}\right)^{11/6} \,.
\ee
 This is the combined SNR from LISA's A and E channels, assuming that the noise is dominated
 by WD confusion noise with (sky-averaged) spectral density $S_n(f) =  S f^{-7/3}$, with 
 $S = 1.44 \times 10^{-44} {\rm Hz}^{4/3}$.
% Note 
%that distance out to which a binary can be detected depends strongly on frequency. 
Let  $\rho_{\rm th}$ the signal-to-noise threshold, such that GWDBs with SNR $ > \rho_{th}$
are immediately subtracted from the data (or otherwise accounted for) {\it before} searching for
(high-z, weaker) MBHBs or EMRIs.  We shall take $\rho_{th} = 50$ as a reasonable fiducial value.
\be\label{Dmin}
%D_{\rm min}(f_i) = \left(\frac{4.31}{\rho_{\rm th}}\right) \left(\frac{\cal M}{0.5224 M_{\odot}}\right)^{5/3} 
D_{\rm min}(f_i) = 2.5\times 10^{-2}\left(\frac{50}{\rho_{\rm th}}\right) \left(\frac{\cal M}{0.25 M_{\odot}}\right)^{5/3} 
\left(\frac{T_{obs}}{3 {\rm yr}}\right)^{1/2}\left(\frac{f_i}{1 {\rm mHz}}\right)^{11/6} \,\, {\rm kpc}.
\ee         

It is absolutely crucial that there be
{\it some} such threshold.  Since $x_i \propto 1/D_i$, if there were no threshold we would have
$p(x_i) \propto x_i^{-3}$ at large $x_i$, and therefore  $P_N(X)$ would fall off only as  $X^{-3}$
at large $X$.    We believe the cut-off is physically reasonable, since there is no reason one
cannot subtract off the very bright sources before looking for weaker ones. (Of course, at the
very end of the data analysis one will want to find the joint best fit for all sources, which will
involve re-adjusting the parameters of all the sources, including the ones that were initially "subtracted out".)

Note, however, that we are imagining removing {\it only} the very strongest GWDBs in the chosen band.  
Now, we expect that LISA data analysis will actually proceed in stages,
and our GWDB model basically represents this background at a rather early stage in the analysis.
At a later stage, we expect that that it will be possible to identify and subtract out all GWDBs with frequencies above a few
mHz.  In principle we could certainly adjust the value of $f_{\rm max}$ in Eq.~(\ref{fprob}) for different stages in the
data analysis , but for simplicity in this paper we just adopt one fixed value for  $f_{\rm max}$.  

Finally the distribution (\ref{varphiprob}) for $p(\varphi_i)$ simply states that the initial orbital phase (and hence also the GW phase) of each binary is random and uniformly distributed.   This uniform distribution in initial phase is what leads to  $p(x) = p(-x)$; i.e. negative values of $(h|h_i)$ are just as likely as 
positive ones.

\subsection{Distribtution of binary parameters for EMRIs}

We next turn to unresolved EMRIs as a source of confusion noise.  There are three types of
EMRIs, since the inspiraling compact object can be a  WD, a neutron star (NS), or a BH. 
BC2 estimated the spectral density of confusion noise from each
of these populations. The estimates are uncertain by at least an order of magnitude, due to the
uncertainty in EMRI capture rates. EMRI confusion noise could end up being comparable to LISA's instrumental noise, and perhaps
even comparable to GWDB confusion noise,  for rates at the high end of the estimated range.
The WDs and NSs would at first seem to represent a bigger confusion
problem, since more than $90\%$ of the GW signal from NS and WD EMRIs will come from unresolvable sources; i.e., 
the NS and WD signals mostly represent confusion noise.  BHs are more massive and so give stronger, more
readily resolved signals; perhaps only $30\%$ of the GW signal from all BH EMRIs is unresolvable.
Nevertheless, since mass segregation tends to concentrate the heavier BHs closer to the MBH, and
since supernova kicks may effectively empty the inner few parsecs of NSs, our judgement is that,
of the three source types, BHs are the most likely to lead to substantial confusion noise.  For this reason, and
for simplicity, we consider
%Since confusion noise from BH EMRIs could possibly dominate the LISA noise curve (while this
%seems much less likely for WD or NS EMRIs), in this section we study 
a population model composed entirely of BH EMRIs. 
%For simplicity, we take all the inspiraling BHs to be $10 M_\odot$, but we
%consider an astrophysically reasonable range of MBH masses.
Since the total signal from BH EMRIs will be dominated
by events at cosmological distances, our model takes cosmological effects into account, and we
include the effects of evolution in both the MBH mass and the event rate.
We adopt the following distribution as our population model:  \eject
\bes\label{EMRIprobs}
\bea
p(M_i)dM_i &=& \theta[M_i - 10^5(1+z_i)^{-0.6}M_\odot]\theta[10^7(1+z_i)^{-0.6}M_\odot - M_i]\nn \\
&& \times \frac{17}{24}\left(10^{119/24} - 10^{85/24}\right)^{-1} \left[\frac{M_i}{(1+z_i)^{-0.6} M_\odot}\right]^{17/24}\frac{dM_i}{M_i}, \label{Mprob} \\
p(\mu_i)d\mu_i &=& \theta[\mu_i - 5M_\odot] \theta[15M_\odot-\mu_i]\frac{1}{\ln 3}\frac{d\mu_i}{\mu_i} , \label{muprob}  \\  
p(z_i) dz_i &=& \theta(z_i-z_c)\theta(2-z_i)\, \mathcal{N}(z_c)[H_0D_L(z_i)]^2 \frac{(1+z_i)^{-1.4}}{\sqrt{0.3(1+z_i)^3 + 0.7}}\, d z_i,  \label{zprob} \\
p(\varphi_i)d\varphi_i &=& \frac{d\varphi}{2\pi} , \label{EMRIvarphiprob}\\
p(f_i)df_i &=& \theta(f_i - f_{\rm min})\frac{8}{3f_{\rm min}}\left(\frac{f_i}{f_{\rm min}}\right)^{-11/3}\, df_i.\label{EMRIfprob}
\eea
\ees
Here $M_i$ and $\mu_i$ are the locally-measured masses of the MBH and stellar-mass BH, respectively.
The mass probability distributions (\ref{Mprob}) and (\ref{muprob}) are derived from the following considerations. The actual distribution $p(\mu_i)$ is very poorly known, but seems centered
on $\mu_i \approx 10 M_\odot$, so we simply assume a flat distribution between $5$ and $15 M_\odot$.
For $p(M_i)$ we restrict attention to the MBHs that {\it today} have masses between $10^{5}$ and
$10^7 M_\odot$, and we assume that their masses have been increasing in time
%The redshift dependence of the minimum and maximum locally measured masses in (\ref{Mprob}) comes from assuming that the locally measured massive black hole masses increase in time 
like $t^{1/2} \propto (1+z)^{-0.6}$, as they continuously swallow gas and compact objects. The locally measured distribution of compact object masses is assumed independent of time. The dependence of probability distributions (\ref{Mprob}) and (\ref{muprob}) on $M$ and $\mu$ respectively is obtained from assumptions on the scaling of merger rates with masses. Let $N$ be the number of mergers with masses comprised between $M$ and $M +dM$ and $\mu$ and $\mu +d\mu$. The rate $R$ of mergers within this box of mass parameters is 
\be
R = \frac{dN}{d f}\dot{f}.
\ee
Following Gair et al.~\cite{Gairetal}, we take the rate $R$ to be proportional to $M^{3/8}$. Since we are considering only a rather narrow range (a factor of 3) of 
masses for the inspiraling object (and since the distribution of stellar BH masses is poorly known), we approximate $R$ as being independent of $\mu$.
These assumptions lead to the following scaling relation
\be
\frac{dN}{df} \propto \frac{R}{\dot{f}} \propto \frac{M^{3/8}}{f^{11/3} \mu M^{2/3}} = f^{-11/3}\mu^{-1}M^{-7/24}.
\ee
From the definition of $N$ we immediately obtain $p(M) \propto M^{-7/24}$ and $p(\mu) \propto \mu^{-1}$. Note incidentally that this also gives the probability distribution (\ref{EMRIfprob}), which was previously derived using the fact that the probability of finding a binary between frequency $f$ and $f + df$ is proportional to the fraction of the binary lifetime it spends around that frequency.

We restrict attention to sources at redshift $z \le 2$, partly since the rates at higher redshift are even more highly uncertain, and
partly since the summed contribution from the $z>2$ sources, all individually weak, clearly will be much more nearly
Gaussian than the noise from the $z<2$ population.

%The redshift probability distribution assumes that all relevant EMRIs are distributed within the spherical volume $z \leq 2$, sources located further away being mostly redshifted below the LISA band. 
The distribution (\ref{zprob})  for $p(z_i)$ 
is then obtained directly from Eq.(10) of Ref.\cite{CurtBBO}, assuming that the locally measured capture rate $\dot{n}$ scales as $\dot{n} \propto (1+z)^{0.6}$, i.e. the capture rate decreases over time as $t^{-1/2}$. This decrease reflects the fact that the MBH first swallows the closest objects, and then it has to wait longer and longer for further compact objects to diffuse in~\cite{Gairetal}.  As with the case of galactic white dwarf binaries, we impose a short-distance cutoff $z_c$, reflecting the fact that very nearby sources can be easily identified and taken out of the confusion noise. In this paper we adopt the nominal value $z_c = 0.1$.  BH EMRIs closer than $z_c$ would typically have matched-filter SNRs  in excess of $300$~\cite{Gairetal}, and so should be very easily identified early in the data analysis. The normalization constant $\mathcal{N}(z_c)$ (defined in Eq.~\ref{zprob}) associated with this 
choice is $\mathcal{N}(0.1) = 1.03044$.   

\section{Applications to various searches}\label{sec:apps}

In this section we apply the statistical tools of section \ref{sec:foundations} to the cases of matched-filter searches for MBHBs or EMRIs buried in confusion noise.  As emphasized in the Introduction, we focus on the weakest resolvable signals of these types. We want to 
assess the importance of the non-Gaussian tails of the SNR distribution in setting the appropriate detection thresholds for these searches.

\subsection{MBHB search: confusion from galactic white dwarf binaries}\label{initMBHB-WD}

We first consider the problem of searching for  MBHB signals buried in the confusion noise from GWDBs. 
As a particular MBHB template signal chirps upwards in frequency, its track on the t-f plane intersects the tracks of
all the white-dwarf binaries in the galaxy (up to the final merger frequency of the MBHB). 
The GWDBs have random parameters, with PDF given by Eqs.~(\ref{Dprob})-(\ref{varphiprob}),  so each GWDB contributes
some amount $x_i$ to the detection statistic.   

Our first goal is to obtain the PDF $p(x_i)$, from which we will
estimate $P_N(X)$ using the Edgeworth expansion.
Our parent variable $x_i$ is an individual overlap given by

\be
x_i =  \frac{1}{S_n[f_i(t_i)]} A_i(t_i) A(t_i) |\delta \dot{f}_{i}(t_i)|^{-1/2} \cos[\delta \Phi_{i} + \text{sgn}(\delta \dot{f}_{i}) \pi/4]
\ee

An important point to notice here is that since our template is a MBHB, it is chirping much faster than any GWDB. Thus to a good approximation we have
\be\label{noWDchirp}
|\delta\dot{f}_i(t_i)| = \dot{f}(t_i).
\ee
In fact, the white dwarf binaries are chirping so slowly that it is reasonable to assume that their frequencies are constant for our analysis.

Eq.~(\ref{Edgeworthexplicit}) requires the first three non-trivial cumulants of the parent variable $x_i$, which are

\bes
\bea  
\kappa_2 &=& E[x_i^2] = \sigma_x^2 \\
\kappa_3 &=& E[x_i^3] \\
\kappa_4 &=& E[x_i^4] - 3 \sigma_x^4
\eea
\ees
Clearly, since the random phase $\delta \Phi_i$ is uniformly distributed between $0$ and $2\pi$, all odd cumulants of $x_i$ vanish and so $\kappa_3 = 0$. We now compute $\kappa_2$ and $\kappa_4$. For the second cumulant, we have
\bea
\kappa_2 &=& \int x_i^2 \, p(D_i)dD_i \, p(f_i)df_i \, p(\varphi_i)d\varphi_i  \nn \\
&=&  \int p(D_i)dD_i \, p(f_i)df_i \frac{A_i^2(t_i) A^2(t_i) }{2S_n^2[f_i(t_i)]} \frac{1}{\dot{f}(t_i)}\label{kappa2_I}
\eea
To go further we need a specific expression for $S_n(f)$.  For simplicity we will approximate $S_n(f)$ in the band $[f_{\rm min},f_{\rm max}]$ (with $f_{\rm min} \sim 10^{-4}\, {\rm Hz}$ and $f_{\rm max} \sim 10^{-2}\, {\rm Hz}$) by the spectral density of the GWDB background (which does indeed work well
throughout most of this band), {\it including} the contribution from GWDBs with $f > 3$ mHz (which should be resolvable at a later stage of the data
analysis).  LISA's instrumental noise rises steeply below and above this band, so we approximate $1/S_n(f)$ as vanishing outside it.
Following \cite{BC2}, we therefore approximate $1/S_n(f)$ by
%\be
%S_n(f) = \left\{ \begin{array}{l}  \infty \,,\,\, f < f_{\rm min} \\ 1.44\times 10^{-44} \left( f / 1 \, {\rm Hz}\right)^{-7/3} \, {\rm Hz}^{-1} \,, \,\, f_{\rm min} \leq f \leq f_{\rm {max}} \\ \infty \,,\,\, f > f_{\rm max} \end{array} \right.
%\ee
%This implies
\be\label{Sninverse}
\frac{1}{S_n(f)} = \theta(f-f_{\rm min})\theta(f_{\rm max}-f)\frac{f^{7/3}}{S},
\ee
with $S = 1.44\times 10^{-44} \,{\rm Hz}^{4/3}$. This also implies $I = (f_{\rm max} - f_{\rm min}) / S$ (cf. Eq.~\ref{Idef}). Combining (\ref{Aidef}), (\ref{templatedef}), (\ref{Dmin}), (\ref{noWDchirp}) and (\ref{Sninverse}), we obtain the following general formula for all even raw moments: 
\be
E[x_i^{2n}] = \frac{(2n-1)!!}{(2n)!!}\left[\frac{A_0^2\mathcal{M}_i^{10/3}}{4\pi IS^2}\right]^n\int_{f_{\rm min}}^{f_{\rm max}}p(f_i) f_i^{11n/3} df_i \int_{D_{\rm min}(f_i)}^{D_{\rm max}} p(D_i) D_i^{-2n} dD_i.
\ee
Note that the lower bound in the frequency integral assumes that the initial template frequency $f_0$ lies below the LISA band lower bound $f_{\rm min}$. If one is interested in templates that begin inside the LISA band, then the lower bound on the frequency integral should be replaced by $f_0$.
%since in that case the template overlaps efficiently only with white dwarf binary signals of frequency higher than $f$. 

In evaluating the distance integrals, we shall assume here that $D_{\rm max} \gg D_{\rm min}(f_i)$ for $f_{\rm min} < f_i < f_{\rm max}$. We can then  approximate $p(D_i)$ by
\be
p(D_i) \simeq \theta(D_i-D_{\rm min})\theta(D_{\rm max}-D_i)\frac{2}{D_{\rm max}^2}D_i,
\ee
which simplifies results considerably\footnote{For a signal-to-noise threshold $\rho_{\rm th} \sim 50$ and three years of observation, the ratio $D_{\rm min}^2 / D_{\rm max}^2$ is of order $10^{-9}$ at $0.1\,{\rm mHz}$, $\sim 10^{-5}$ at $1\, {\rm mHz}$ and $\sim 0.03$ at $10 \, {\rm mHz}$. Thus the
error from this approximation is negligible throughout the band $0.1-10$mHz.}.
The computation of the remaining integrals are straightforward and the results are
\bes
\bea
E[x_i^2] &=& \frac{1}{2}\left[\frac{A_0^2\mathcal{M}_i^{10/3}}{4\pi IS^2}\right]\left[\frac{16f_{\rm min}^{8/3}}{3D_{\rm max}^2}\right]\Bigg(f_{\rm max}\left\{\ln\left[\frac{D_{\rm max}}{D_{\rm min}(f_{\rm max})}\right] +\frac{11}{6}\right\} - f_{\rm min}\left\{\ln\left[\frac{D_{\rm max}}{D_{\rm min}(f_{\rm min})}\right] +\frac{11}{6}\right\}\Bigg) \\
E[x_i^{2n}] &=& \frac{(2n-1)!!}{(2n)!!}\left[\frac{A_0^2\mathcal{M}_i^{10/3}}{4\pi IS^2}\right]^n\left[\frac{16f_{\rm min}^{8/3}}{3D_{\rm max}^2}\right]\left[\frac{(f_{\rm max} - f_{\rm min})(1\, {\rm mHz})^{11(n-1)/3}}{2(n-1)D_{\rm min}^{2(n-1)}(1\, {\rm mHz})}\right] \,\, {\rm for} \, n \geq 2. 
\eea
\ees

We then obtain the following expression for the fourth cumulant

\bea
\kappa_4 &=& \sigma_x^4\left\{\frac{9\bar{f}_{\rm min}^{-8/3}(\bar{f}_{\rm max} - \bar{f}_{\rm min})}{64}\left[\frac{D_{\rm max}}{D_{\rm min}(1\, {\rm mHz})}\right]^2 
\right. \nn \\ && \times  \left.
\Bigg(\bar{f}_{\rm max}\left\{\ln\left[\frac{D_{\rm max}}{D_{\rm min}(f_{\rm max})}\right] +\frac{11}{6}\right\} - \bar{f}_{\rm min}\left\{\ln\left[\frac{D_{\rm max}}{D_{\rm min}(f_{\rm min})}\right] +\frac{11}{6}\right\}\Bigg)^{-2}  -  3\right\}, \label{kappa4_I}
\eea
where $\bar{f} = f / (1\,{\rm mHz})$. Since $\bar{f}_{\rm min}\sim  0.1$ and $\Delta^2 \equiv D^2_{\rm max} / D_{\rm min}^2(1 \,{\rm mHz}) = (10\rho_{\rm th}/1.26)^2 \sim 10^5$, the first term in (\ref{kappa4_I}) is clearly much larger than the second and so we can safely drop the $-3$ term. Substituting the resulting fourth cumulant into (\ref{Edgeworthexplicit}) finally yields the following PDF for the signal-to-noise ratio

\bea
P_N(X) &=& \frac{1}{\sqrt{2\pi N\sigma_x^2}}e^{-\frac{X^2}{2N\sigma_x^2}}\left[1 + \frac{3\bar{f}_{\rm min}^{-8/3}(\bar{f}_{\rm max} - \bar{f}_{\rm min})\Delta^2}{512N} 
\right. \nn \\ && \times \left.
\Bigg\{(\bar{f}_{\rm max} - \bar{f}_{\rm min})\log\Delta + \frac{11}{6}\Big[\bar{f}_{\rm max}(1+\log\bar{f}_{\rm max}) - \bar{f}_{\rm min}(1+\log\bar{f}_{\rm min})\Big]\Bigg\}^{-2} H_4\left(\frac{X}{\sqrt{N\sigma_x^2}}\right)\right] \label{answer}
\eea
In the above the number $N$ is the number of galactic white dwarf binaries with gravitational wave frequency above $f_{\rm min}$, our assumed lower bound of the LISA band\footnote{Again, if one is interested in a template which begins inside the LISA band at $t=0$, then the number $N$ appearing in (\ref{answer}) is the number of galactic white dwarf binaries in the frequency interval $(f_0,f_{\rm max})$, with $f_0$ being the initial template frequency.}. Using %LISA band limits of 
$\bar{f}_{\rm min} \sim 0.1$,  $\bar{f}_{\rm max} \sim 10$, a threshold $\rho_{\rm th} \sim 50$, 3 years of observation and assuming $N = 3\times10^7$ galactic binaries contributing to the confusion noise in the LISA band, the relative size of the $1/N$ correction to the Gaussian PDF predicted by the Central Limit Theorem at the $7\, \sigma$ is $\sim 0.02$. Thus, at the $7\, \sigma$ level, the Edgeworth analysis
shows that the non-Gaussian tails of the SNR distribution is negligible for this search.

Finally, notice that if we take the inner cut-off distance $D_{\rm min}(f)$ as fixed, then the
result (\ref{answer}) is actually independent of our overall factor $A_0$ in the waveform amplitudes,
and also independent of the magnitude of $S$ in Eq.~(\ref{Sninverse}) for the noise spectral density appearing
in the inner product.  Multiplying either of these by an overall factor simply re-scales 
all the $x_i$, $X$, and $\sigma_x$  by the same amount, while the result (\ref{answer}) is expressed
purely in terms of the dimensionless ratio $X/\sigma_x$.

\subsection{EMRI search: confusion from galactic white dwarf binaries}

Notice here the important fact that our results in IV.B  are independent of the chirp mass $M_c$ of  the normalized search template. The reason for this is that while the template amplitude (for normalized templates) scales like $M_c^{5/6}$ [cf. Eq.(\ref{templatedef})], the time over which there is significant overlap
with any GWDB signal scales like $(\dot f)^{-1/2} \propto M_c^{-5/6}$.
%
%while the overlap gets weaker [cf. Eqs.(\ref{noWDchirp}) and (\ref{omegaorbdot})] at a rate that precisely compensates the increase in the template amplitude. 
Therefore  the analysis in  IV.A applies with practically no modification to searches for EMRIs buried in galactic white dwarf confusion noise. In the case of a realistic EMRI search in Gaussian noise, the detection threshold is around $\sim 14 \, \sigma$. At that level the relative correction to $P_N(X)$ predicted by the $1/N$ term in the Edgeworth series is  $\sim 0.4$.  Since this correction is of order unity, one should also check the $1/N^2$ term.

Performing that calculation using the results of this section and the next-to-leading term of Edgeworth series (given by, e.g., Petrov~\cite{Petrov}) yields a next-to-leading correction of order $\sim 0.2$. Since this correction term is smaller than the leading term, we are inclined to trust the leading order correction to the Gaussian distribution.  This confirms that the Gaussian approximation is still reasonably accurate at the $14\,\sigma$ level, 
for EMRIs buried in galactic white dwarf confusion noise.
% may also be treated as Gaussian for an EMRI search.
   
\subsection{MBHB search: confusion from EMRIs}

In this subsection we compute the signal-to-noise PDF for a matched-filter search for MBHB signals buried in confusion noise from unresolved EMRIs. At any given time it is expected that $\sim 10^{4-6}$ unresolved EMRIs signals will be radiating GWs into the band $0.1-10$ mHz, representing a significant
source of confusion noise. When searching for MBHBs in a background of EMRIs, we may again assume that the parameter $\delta \dot{f}_i$ appearing in stationary phase overlap (\ref{analyticoverlap}) is entirely dominated by the chirping MBHB. However, since the radiation reaction timescale for EMRIs is comparable to the LISA mission lifetime, our calculation must take into account that some EMRIs
that are "live" (i.e., are pre-merger) at the beginning of LISA's observation period will "die" (merge)
before their t-f track can be crossed by the MBHB's t-f track.  

\subsubsection{Edgeworth expansion}

As before, we wish to compute the cumulants of the parent distribution $p(x_i)$ to obtain the Edgeworth expansion of $P_N(X)$. Following subsection \ref{initMBHB-WD}, we first derive an expression for the raw moments of the parent distribution, i.e.

\bea
E[x^{2n}] &=& \frac{(2n-1)!!}{(2n)!!}\int p(z_i) \, dz_i \,\, p(f_i)\,df_i \,\, p(M_i)\, dM_i\,\, p(\mu_i)\,d\mu_i \frac{A_i^{2n}(t_i) A^{2n}(t_i)}{S^{2n}_n[f(t_i)]}\frac{1}{|\dot{f}(t_i)|^n} \nn \\
&=& \frac{(2n-1)!!}{(2n)!!}\left[\frac{A_0^2}{4\pi I S^2}\right]^n\int \frac{(1+z)^{10n/3}}{D_L^{2n}(z_i)}p(z_i) dz_i \int p(M_i)dM_i \, p(\mu_i) d\mu_i \, p(f_i)df_i \,M_i^{4n/3}\mu_i^{2n}f(t_i)^{11n/3} 
\nn \\ && \times
\theta[f(t_i) - f_{\rm min}]\theta[f_{\rm max}-f(t_i)]\theta(t_i).
\eea
The step function $\theta(t_i)$ sets to zero the contribution from EMRIs whose t-f tracks do not
cross that of the MBHM template within the band $[f_{\rm min}, f_{\rm max}]$ (since if $t_i < 0$, the
tracks must cross at some frequency below $f_{\rm min}$).
%is needed to restrict the integral to the region of parameter space such that the $t-f$ tracks cross at some time $t_i > 0$.
In terms of the following parameters
\bes
\bea
\alpha_i &=& \left(\frac{f_i}{f_0}\right)^{8/3}, \\
\beta_i &=& \frac{\mu M^{2/3}}{\mu_i M^{2/3}_i},
\eea
\ees
the crossing time $t_i$ is given by

\be
t_i = \frac{t_{\rm rr}}{\alpha_i}\left(\frac{\alpha_i - 1}{1 - \beta_i^{-1}}\right),
\ee
which then gives

\be\label{1mtovertrr}
1 - \frac{t_i}{t_{\rm rr}} = \frac{1}{\alpha_i} \left(\frac{\beta_i - \alpha_i}{\beta_i - 1}\right).
\ee
%The above expression must be positive for a consistent solution to exist. 
As before, we shall assume that $f_0 < f_{\rm min}$, i.e. the MBHB template begins outside the LISA band, but still at high-enough frequency that it has time to sweep through the LISA band during the mission lifetime. Then $\alpha_i > 1$, and 
combining this with the condition $t_i \geq 0$ implies that $\beta_i > \alpha_i$.

Switching integration variables from $(f_i,\mu_i)$ at fixed $M_i$ to $(\alpha_i,\beta_i)$, we then obtain
\bea
E[x^{2n}] &=& \frac{(2n-1)!!}{(2n)!!}\left[\frac{A_0^2\mu^2M^{4/3}f^{11/3}}{4\pi I S^2}\right]^n\int \frac{(1+z)^{10n/3}}{D_L^{2n}(z_i)}p(z_i) dz_i 
\int p(M_i)dM_i \int_{\beta_{\rm min}}^{\beta_{\rm max}} \frac{d\beta_i}{\beta_i \ln 3} \int_{\alpha_{\rm min}}^\infty \alpha_{\rm min}\frac{d\alpha_i}{\alpha_i^2}
\nn \\ && \times
{\beta_i^{-2n}} \alpha_i^{11n/8}\left(\frac{\beta_i-\alpha_i}{\beta_i - 1}\right)^{-11n/8}
%\theta[f(t_i) - f_{\rm min}]\theta[f_{\rm max}-f(t_i)]\theta(\alpha_i - 1)\theta(\beta_i-\alpha_i),
\theta[f(t_i) - f_{\rm min}]\theta[f_{\rm max}-f(t_i)]\theta(\beta_i-\alpha_i),
\eea
%where $\alpha_{\rm th} = (f_{\rm min} / f)^{8/3}$. 
where $\alpha_{\rm min} \simeq (f_{\rm min} / f_0)^{8/3}$, the exact expression given below in (\ref{alphaminmax}). 

Using (\ref{1mtovertrr}), we can rewrite the step functions as follows
\be
\theta[f(t_i) - f_{\rm min,max}] = \theta[\alpha_i - \alpha_{\rm min,max}],
\ee
where
\be\label{alphaminmax}
\alpha_{\rm min,max} = \left(\frac{f_{\rm min,max}}{f_0}\right)^{8/3}\left[1 - \frac{1}{\beta_i} + \frac{1}{\beta_i}\left(\frac{f_{\rm min,max}}{f_0}\right)^{8/3}\right]^{-1}.
\ee
We then get 

\bea
E[x^{2n}] &=& \frac{(2n-1)!!}{(2n)!!}\left[\frac{A_0^2\mu^2M^{4/3}f^{11/3}}{4\pi I S^2}\right]^n\int \frac{(1+z)^{10n/3}}{D_L^{2n}(z_i)}p(z_i) dz_i \int p(M_i)dM_i \int_{\beta_{\rm min}}^{\beta_{\rm max}} \frac{d\beta_i}{\beta_i \ln 3} 
\nn \\ && \times
%\int_{{\rm max}[\alpha_{\rm min},1]}^{{\rm min}[\beta_i,\alpha_{\rm max}]} \alpha_{\rm th}\frac{d\alpha_i}{\alpha_i^2} {\beta_i^{-2n}} \alpha_i^{11n/8}\left(\frac{\beta_i-\alpha_i}{\beta_i - 1}\right)^{-11n/8}.
\int_{\alpha_{\rm min}}^{\alpha_{\rm max}} \alpha_{\rm min}\frac{d\alpha_i}{\alpha_i^2} {\beta_i^{-2n}} \alpha_i^{11n/8}\left(\frac{\beta_i-\alpha_i}{\beta_i - 1}\right)^{-11n/8}.
\eea

Strictly speaking the upper integration limit over $\alpha_i$ should be ${\rm min}(\alpha_{\rm max},\beta_i)$. However we show below in (\ref{upperlim}) that $\beta_i > \alpha_{\rm max}$, which explains justifies our limit in the previous equation. 
Performing the $\alpha_i$ integral yields

\bea
E[x^{2n}] &=& \frac{(2n-1)!!}{(2n)!!}\left[\frac{A_0^2\mu^2M^{4/3}f^{11/3}}{4\pi I S^2}\right]^n\frac{\alpha_{\rm th}}{\ln 3}\int \frac{(1+z)^{10n/3}}{D_L^{2n}(z_i)}p(z_i) dz_i \int p(M_i)dM_i 
\nn \\ && \times
\int_{\beta_{\rm min}}^{\beta_{\rm max}} \frac{8(\beta_i - 1)^{11n/8}}{(11n-8)\beta_i^2}\left[\alpha_i^{11n/8-1}(\beta_i - \alpha_i)^{-11n/8 + 1}\right]_{\alpha_{\rm min}}^{\alpha_{\rm max}}d\beta_i. \label{Ex2ntemp}
\eea
Now from (\ref{alphaminmax}), we have

\be\label{upperlim}
\beta_i - \alpha_{\rm min,max} = \left(\frac{f_{\rm min,max}}{f_0}\right)^{-8/3}(\beta_i - 1)\alpha_{\rm min,max},
\ee
which, when substituted into (\ref{Ex2ntemp}), yields the following 

\bea
E[x^{2n}] &=& \frac{(2n-1)!!}{(2n)!!}\left[\frac{A_0^2\mu^2M^{4/3}f^{11/3}}{4\pi I S^2}\right]^n\frac{\alpha_{\rm th}}{\ln 3}\int \frac{(1+z)^{10n/3}}{D_L^{2n}(z_i)}p(z_i) dz_i \int p(M_i)dM_i 
\nn \\ && \times
\int_{\beta_{\rm min}}^{\beta_{\rm max}} \frac{8(\beta_i-1)}{(11n - 8)\beta_i^2}\left[\left(\frac{f_{\rm max}}{f_0}\right)^{(11n-8)/3} - \left(\frac{f_{\rm min}}{f_0}\right)^{(11n-8)/3}\right]d\beta_i. \label{Ex2ntempII}
\eea
Now since $\beta_i \sim 10^5$ over the integration range $(\beta_{\rm min},\beta_{\rm max})$, we may set $\beta_i - 1 \simeq \beta_i$ to very good accuracy and perform the remaining $\beta_i$ integral, which simply gives $\ln 3$, the normalization constant of the $\beta_i$ probability distribution. Then the $M_i$ integral trivially gives unity, as the remaining integrand, apart from $p(M_i)$, is independent of $M_i$. We are then left with

\be\label{rawmoments}
E[x^{2n}] = \frac{(2n-1)!!}{(2n)!!}\frac{8\alpha_{\rm th}}{(11n - 8)}\left[\frac{A_0^2\mu^2M^{4/3}f^{11/3}}{4\pi I S^2}\right]^n\left\{\left[\frac{f_{\rm max}}{f_0}\right]^{(11n-8)/3} - \left[\frac{f_{\rm min}}{f_0}\right]^{(11n-8)/3}\right\}\int \frac{(1+z)^{10n/3}}{D_L^{2n}(z_i)}p(z_i) dz_i.
\ee
The redshift integrals must be performed numerically and we denote each value as

\be
\int_{z_{\rm min}}^{z_{\rm max}} \frac{(1+z)^{10n/3}}{D_L^{2n}(z_i)}p(z_i) dz_i \equiv H_0^{2n}\,\zeta_n.
\ee
For inner cutoff redshift $z_{\rm min} = 0.1$ and $z_{\rm max} = 2$, the first few values of $\zeta_n$ are 

\bes\label{Zvals}
\bea
\zeta_1 &=& 4.17924 \\
\zeta_2 &=& 29.72232 \\
\zeta_3 &=& 763.5828
\eea
\ees

The first two terms of the Edgeworth-expanded PDF for the signal-to-noise ratio are then given by

\be 
P_N(X) = \frac{1}{\sqrt{2\pi N\sigma_x^2}}e^{-\frac{X^2}{2N\sigma_x^2}}\left\{1 + \frac{9}{1792\, N}\left[\frac{(1 - x^{14/3})}{x^{8/3}(1-x)^2}\frac{\zeta_2}{\zeta_1^2}\right] H_4\left(\frac{X}{\sqrt{N\sigma_x^2}}\right) + O(N^{-2})\right\} \label{answerII},
\ee
where  $N$ is the number of EMRIs that lie in the LISA band and where $x \equiv f_{\rm min} / f_{\rm max}$. Taking $N = 5\times 10^5$, the same frequency limits for the LISA band as before and using (\ref{Zvals}), the relative size of the $1/N$ correction to the Gaussian PDF predicted by the Central Limit Theorem at the $7\,\sigma$ level (i.e., at $X/\sqrt{N\sigma_x^2} = 7$) is found to be $\approx  8$. Since this "first-order correction" is already larger than the zeroth-order estimate, the Edgeworth expansion simply  cannot provide a reliable answer for this problem. 
Instead,  the problem of searching for MBHBs buried in EMRI confusion noise must be addressed within the context of the theory of large deviations, to which we turn next.

\subsubsection{Large-deviations analysis}

Here we compute the signal-to-noise PDF for a search or MBHBs buried in EMRI confusion noise, following the prescription of large-deviations theory. The starting point is the construction of the modified cumulant generating functional $\lambda(\beta)$. Since we have already computed analytically all the raw moments of the parent distribution, we may evaluate $\lambda(\beta)$ from its power series expansion numerically to any desired accuracy, i.e. we use 

\bea
e^{\lambda(\beta)} &=& \sum_{p=0}^\infty \frac{\beta^p}{p!}E[x^p] \nn \\
&=& 1 + \frac{1}{2}(\beta\,\sigma_x)^2 + \sum_{n=2}^\infty \frac{(\beta \, \sigma_x)^{2n}}{(2n)!} \frac{E[x^{2n}]}{\sigma_x^{2n}}
\eea
Defining $\tilde{\beta} = \beta \sigma_x$ and using (\ref{rawmoments}), we obtain the following expression for the cumulant generating functional

\be\label{cumulgenexp}
e^{\lambda(\beta)} = 1 + \frac{1}{2}\tilde{\beta}^2 + 8x^{8/3}\sum_{n=2}^\infty \frac{\bar{\beta}^{2n}}{n!^2}\left[\frac{1-x^{(11n-8)/3}}{11n-8}\right]\zeta_n,
\ee
where

\be
\bar{\beta} = \left[\frac{3}{16x^{8/3}(1-x)\zeta_1}\right]^{1/2}\tilde{\beta}.
\ee
We then compute the rate function describing the signal-to-noise PDF following these steps. First we compute the cumulant generating functional $\lambda$ from (\ref{cumulgenexp}) numerically. The infinite sum is truncated when the $n^{\rm th}$ term of the sum is of order $10^{-10}$ of the sum of the previous $n-1$ terms. Because the sum converges\footnote{This is a consequence of the fact that the parent probability distribution has compact support in our model.} for any value of $\beta$, we are confident that this is a reasonable accuracy criterion. We next compute $\lambda^\prime \equiv d\lambda / d\tilde{\beta}$ (here $\lambda$ is considered an implicit function of $\tilde{\beta}$) by taking a derivative of (\ref{cumulgenexp}) and evaluating the sum numerically using the same truncation criterion as before. The next step is the maximization over $\beta$ (or equivalently $\tilde{\beta}$) of the quantity $I = Z\beta - \lambda = \tilde{Z}\tilde{\beta} - \lambda$, where $\tilde{Z} = 
 Z/\sigma_x$. The value of $\tilde{\beta}$ which maximizes $I$ is simply the one satisfying $\tilde{Z} = \lambda^\prime(\tilde{\beta})$. By inverting numerically the function $\lambda^\prime$, we obtain the function $\tilde{\beta}(\tilde{Z})$. The rate function can then be computed numerically for any desired value of $\tilde{Z}$ as follows

\be
%I(\tilde{Z}) = \tilde{Z}\tilde{\beta}(\tilde{Z}) - \tilde{\lambda}[\tilde{\beta}(\tilde{Z})/\sigma_x].
I(\tilde{Z}) = \tilde{Z}\tilde{\beta}(\tilde{Z}) - \lambda[\tilde{\beta}(\tilde{Z})].
\ee

\begin{figure}
\begin{center}
\epsfig{file=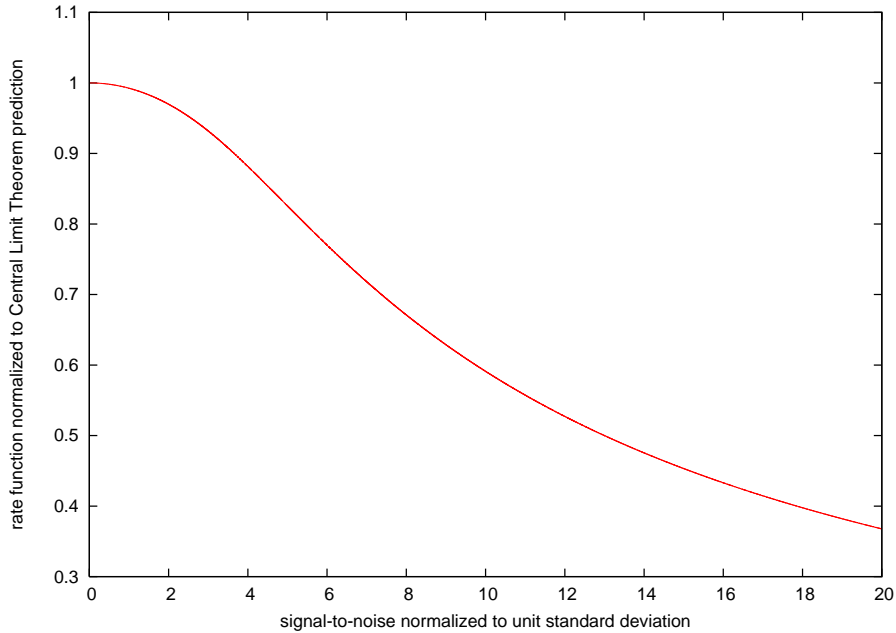,width=8.5cm,angle=270}
\caption{This figure shows the rate function of the signal-to-noise PDF for the MBHB search in EMRI confusion noise for a simple Newtonian chirp toy model. The rate function is normalized to the value predicted by the Central Limit Theorem, namely $I_{CLT} = \tilde{Z}^2 / 2$. This plot was generated using $N = 5\times 10^5$ for the number of unresolved EMRIs in the the band $0.1 - 10\, {\rm mHz}$. If one wants to vary the number of unresolved EMRIs, one simply rescales $x$-values by $\sqrt{N/(5\times 10^5)}$, since this is the multiplicative factor needed to rescale $\tilde{Z}$ to unit standard deviation.}\label{ratefunction}
\end{center}
\end{figure}

The resulting rate function 
of the signal-to-noise PDF for the MBHB search in EMRI confusion noise is plotted in 
Fig.\ref{ratefunction}.  The vertical axis is the actual rate function normalized by the
Central Limit Theorem estimate: $I(Z)/[-0.5 Z^2/\sigma^2_x]$. The horizontal axis is 
the SNR normalized to unit standard deviation, or $N^{1/2} Z/\sigma_x$. The result is
plotted for $N= 5 \times 10^5$ in-band EMRIs, but to obtain the curve for any other value of $N$, one
simply re-scales the x labels on the horizontal axis by $\sqrt{N/(5\times 10^5)}$.

From that figure, one can easily see that at the $7\, \sigma$ level, the rate function derived from large-deviations theory differs significantly from the Central Limit Theorem estimate (confirming our conclusion
from the Edgeworth analysis in IV.C.1).  While the Central Limit Theorem estimate for 
$P_N(X =7 N^{1/2} \sigma_x)$ is $(2\pi)^{-1/2}{\rm exp}[-49/2]$, the actual probability density is
$\approx {\rm exp}[-0.72*49/2]$, or a factor $\sim 10^3$ larger.
Therefore in deciding the appropriate detection threshold, one must take into account the non-Gaussianity of the signal-to-noise PDF.   In the next subsection we discuss the proper adjustment of the detection threshold, based on the rate function of Fig.\ref{ratefunction}.

\subsection{Adjusting the detection threshold}

We have considered searches for MBHB signals buried in two different types of confusion background: GWDBs and EMRIs.
For GWDBs we showed that the PDF for the SNR could be safely approximated as as Gaussian (up to the detection threshold), but that a search
in EMRI confusion noise alone would have to take into account the significant non-Gaussianity in $P_N(X)$ at $X \sim 7 \sigma$.  
However, EMRI confusion noise is
unlikely to dominate the total noise, so in this subsection we show explicitly how to calculate the appropriate detection threshold for noise that is a sum of 
EMRI confusion noise plus Gaussian noise.

Since confusion noise from GWDBs is Gaussian to a good approximation, it can be combined with instrumental noise into one single source of Gaussian noise.  We shall here denote the signal-to-noise ratio obtained by correlating a MBHB template with this Gaussian noise as $\rho_g$. This signal-to-noise is drawn from the following PDF
\be
p_g(\rho_g) = \frac{1}{(2\pi \sigma_g^2)^{1/2}}e^{-\rho_g^2/2\sigma_g^2},
\ee      
where $\sigma_g$ is the standard deviation of the random variable $\rho_g$. Next we denote the signal-to-noise ratio obtained by correlating a MBHB template with EMRI confusion noise as $\rho_c$. This signal-to-noise is drawn from the following PDF

\be\label{pcrhoc}
p_c(\rho_c) = \mathcal{N}_c \exp\left[-\frac{\rho_c^2}{2\sigma_c^2}\tilde{I}\left(\sqrt{\frac{N^\ast}{N}}\frac{\rho_c}{\sigma_c}\right)\right],
\ee
where $\tilde{I}$ is the re-scaled rate function plotted in Fig.\ref{ratefunction}, $N^\ast = 5\times 10^5$ is the number of unresolved EMRIs chosen to generate Fig.\ref{ratefunction}, $\mathcal{N}_c$ is a normalization constant and $\sigma_c$ is the standard deviation of the random variable $\rho_c$. Consider now the PDF for the total signal-to-noise ratio $\rho = \rho_g + \rho_c$. It is given by the following convolution integral 

\be\label{prho}
p(\rho) = \int_{-\infty}^{+\infty} p_g(\rho - \rho_c)p_c(\rho_c)\, d\rho_c,
\ee
which can be performed numerically. If the rate function $\tilde{I}$ were equal to unity, i.e. if $\rho_c$ were Gaussian, then $\rho$ would also be a Gaussian random variable with standard deviation $\sigma = (\sigma_g^2 + \sigma_c^2)^{1/2}$. We shall determine a threshold on the normalized total SNR  $\hat{\rho} \equiv (\sigma_g^2 + \sigma_c^2)^{-1/2}\rho$, assuming that if the rate function $\tilde{I}$ were equal to unity, then the appropriate detection threshold would be set at $\hat{\rho} = 7$. In order words, the acceptable false alarm probability $P_{\rm FA}$ is assumed to be the integral of the Gaussian PDF for $\hat{\rho}$ over the range $(-\infty,-7)$ and $(7,+\infty)$:

\be
P_{\rm FA} = 2 \int_{7}^\infty \frac{1}{\sqrt{2\pi}}e^{-\hat{\rho}/2} d\hat{\rho} = {\rm erfc}(7/\sqrt{2}).
\ee
The actual detection threshold $\hat{\rho}_{\rm th}$ for the MBHB search is then determined by the following equation

\bea
P_{\rm FA} &=& 2\sigma\int_{\hat{\rho}_{\rm th}}^\infty p(\sigma\hat{\rho}) d\hat{\rho} \nn \\
&=& 2\sigma\int_{\hat{\rho}_{\rm th}}^\infty \int_{-\infty}^{+\infty} p_g(\sigma\hat{\rho} - \rho_c)p_c(\rho_c)\, d\rho_c d\hat{\rho} \nn \\
&=& \int_{-\infty}^{+\infty} {\rm erfc}\left[\frac{(1+\varepsilon^2)^{1/2}\hat{\rho}_{\rm th} - \varepsilon\hat{\rho}_c}{\sqrt{2}}\right]p_c(\rho_c) \, d\rho_c, \label{thresheq}
\eea
where $\hat{\rho}_c \equiv \rho_c / \sigma_c$ and where $\varepsilon \equiv \sigma_c / \sigma_g$ measures the relative strength of the non-Gaussian component of the noise.

\begin{figure}
\begin{center}
\epsfig{file=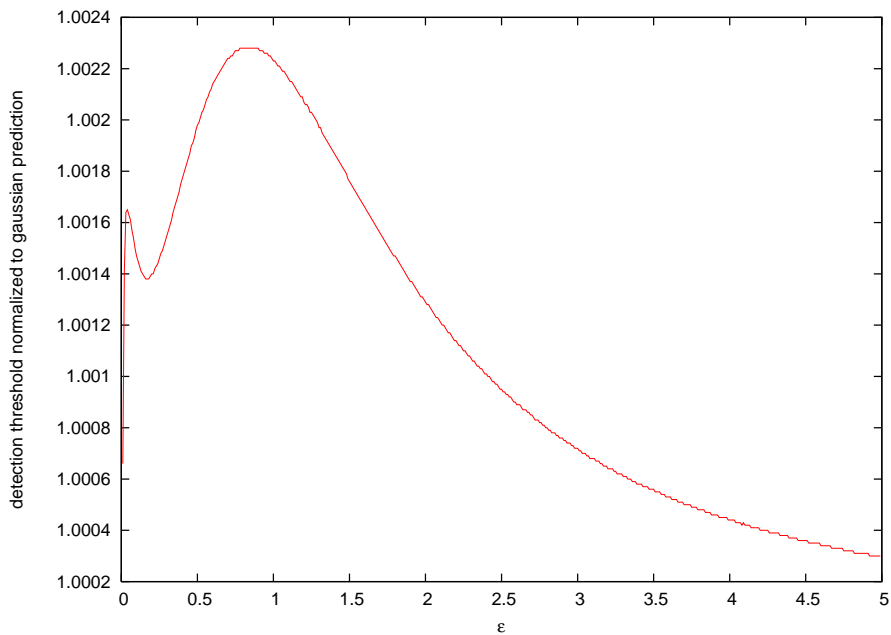,width=8.5cm,angle=270}
\caption{This figure shows the normalized detection threshold $\hat \rho_{th}$ (at fixed false alarm probability) for a total noise composed of a Gaussian component (instrumental noise and GWDB confusion noise) and a non-Gaussian component (EMRI confusion noise) as a function of the ratio $\varepsilon = \sigma_c/\sigma_g$. In this plot we consider the Gaussian component to be fixed and $\varepsilon$ varies by adjusting the number of unresolved EMRIs.   Note that $\hat \rho_{th}$ is 
always nearly one, i.e., nearly the same as for a Gaussian distribution with the same standard deviation.
%This plot is based on the estimate that $\varepsilon = 1$ for $N = 1.25 \times 10^7$ unresolved EMRIs,  so $\varepsilon = \sqrt{N/(1.25\times 10^7)}$.
}
\label{threshold}
\end{center}
\end{figure}

%In Figs.\ \ref{threshold} and \ref{thresholdII} we plot the $\hat{\rho}_{\rm th}$ as function of $\varepsilon$ for two different values of the parameter $\sigma_g$. 
In Fig.\ \ref{threshold}  we plot $\hat{\rho}_{\rm th} / 7$ as function of $\varepsilon$ for our best estimate of  $\sigma_g$ (taken from BC2, assuming
no GWDBs have been fitted out).  That is, we fix the amplitude of the GWDB background and plot how $\hat{\rho}_{\rm th}$ varies as one increases the
number of unresolved in-band EMRIs. This figure was generated as follows.  For any $\varepsilon$ we estimated $N$ (the number of unresolved EMRIs) 
using
\be\label{Neps}
\varepsilon \approx \bigg(\frac{N}{1.25\times 10^{7}}\bigg)^{1/2} \, . 
\ee
(Since an astrophysically reasonable estimate is $N = 5 \times 10^{5}$,   we expect  $\varepsilon \approx 0.2$ in practice.)
We insert $N$ into Eq.~(\ref{pcrhoc}) to  obtain $p_c(\rho_c)$, which  
we than plug into the last line of Eq.~(\ref{thresheq}).
We obtain the detection threshold $\hat \rho_{th}$ by solving (\ref{thresheq}) numerically.  

The most important fact one gleans from
Fig.\ \ref{threshold} 
is that $\hat \rho_{th}$ is always very close to one.  We can understand this as follows.  For realistic values of $N$,
the SNR from EMRIs is significantly non-Gaussian, but since the noise is dominated by instrumental
and GWDB background noise, the non-Gaussianity of the
EMRI background has little effect on the threshold.  When $N$ is large enough that EMRI noise is a large fraction of the total noise, the EMRI confusion
noise is much more Gaussian, so again the threshold is very close to the Gaussian prediction.

\begin{figure}
\begin{center}
\epsfig{file=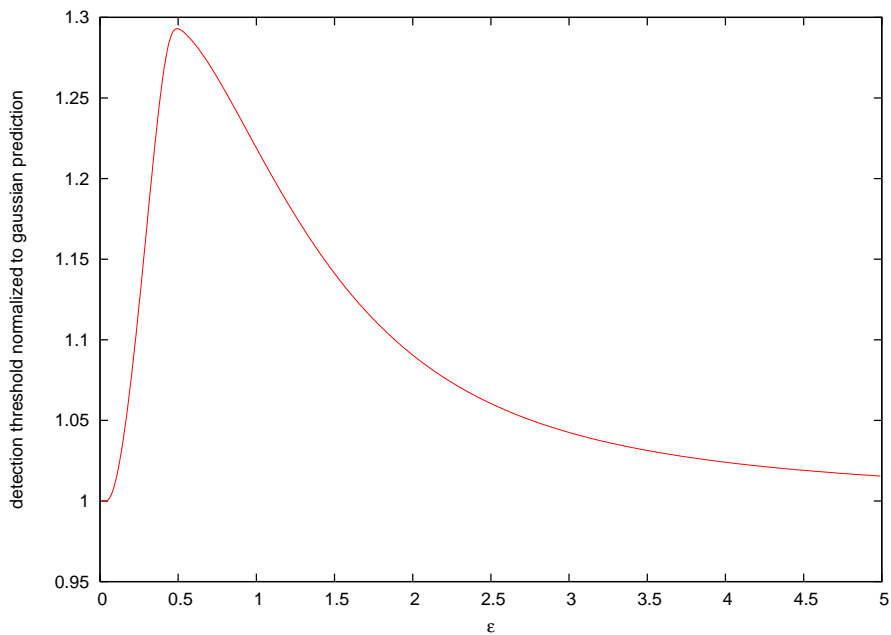,width=8.5cm,angle=270}
\caption{This plot is similar to Fig.\ref{threshold}, but with the strength of the Gaussian component reduced "by hand" by a factor of $50$, i.e. $\sigma_g^2 \rightarrow \sigma_g^2/50$ so that $\varepsilon = \sqrt{N/(2.5\times 10^5)}$.  In this case, the detection threshold  can be up to 
$\sim 30\%$ higher than for a Gaussian distribution with the same standard deviation.}\label{thresholdII}
\end{center}
\end{figure}

Fig.\ \ref{thresholdII}  is the same as Fig.\ \ref{threshold}, except that for illustrative purposes we
have decreased "by hand" the value of $\sigma_g$ by $\sqrt{50}$. 
In this case, the normalized threshold $\hat \rho_{th}$ could be (for $\varepsilon \approx 0.5$) up
to $\sim 1.3$ times higher than for Gaussian noise with the same standard deviation.  For $N = 5 \times 10^5$ and this reduced $\sigma_g$, we would have $\varepsilon  \approx  1.4$ and 
$\hat \rho_{th} \approx 1.15$; i.e., the appropriate threshold would be $8\,\sigma$ instead of $7\,\sigma$.
We note that $(\hat \rho_{th}/7) \rightarrow 1$ both as $\epsilon \rightarrow 0$ and as 
$\epsilon \rightarrow \infty$.  This is easily understood, since as $\epsilon \rightarrow 0$ the
noise becomes just the Gaussian part, while as $\epsilon \rightarrow \infty$ we also
have $N \rightarrow \infty$, so the EMRI portion becomes Gaussian.

\section{Summary, conclusions, and open issues}

In this paper we have analyzed the problem of determining the appropriate detection for several
idealized searches.  The most important simplifications were that we used "lowest-order" waveforms (based on the quadrupole formula, and assuming quasi-circular inspirals) and simplified population distributions for the confusion sources.  
%Crucially, we always imposed a short-distance cut-off on those populations, motivated by the fact that
%the very closest, "brightest" sources should be easily resolved and then removed from the confusion population.  
%Using the Edgeworth expansion, 
We first considered searches for both
MBHB signals and EMRI signals buried in confusion noise from GWDBs.
Using the Edgeworth expansion, we showed that for these cases the PDF of the 
standard detection statistic remains nearly Gaussian out to the relevant detection thresholds.
We then considered searches for MBHB signals buried in just EMRI confusion noise.
In that case, using large-deviations theory, we found that
$7\,\sigma$ events would occur $10^3$ times more often than suggested by the Central Limit
Theorem.  However this third case was rather unrealistic , since it is very unlikely that EMRI confusion
noise will dominate the total LISA noise curve.  We then considered a more realistic example, 
in which  the EMRI confusion noise was combined with Gaussian noise of $\sim 5$ times larger  amplitude. In that case, we again found that the non-Gaussianity of the EMRI confusion noise ends up
having a negligible impact in setting the appropriate detection threshold. 

The rather minimal impact of non-Gaussian tails in these models appears to stem from three circumstances.  First, the number of confusion noise sources is always rather large. Second, in all cases we imposed a short-distance cut-off on the distribution of the
background sources, arguing that the very closest and therefore strongest of the background sources could be
effectively removed (or otherwise taken into account) {\it before} searching for other types of sources. Third, all three
model problems shared the feature that the search templates and background templates evolve in frequency on very different timescales: $\dot f_{WD} \ll \dot f_{EMRI} \ll \dot f_{MBHB}$. Since $x_i \propto |\delta \dot f_i|^{-1/2}$, this
separation of timescales ensures that $p(x)$ has no large outliers arising from coincidentally small $|\delta \dot f_i|^{-1/2}$.
Put another way, the dissimilarity of the searched-for and background signals is crucial to the sharp fall-off of $p(x)$
at large $x$.  The high-$X$ tail of $P_N(X)$ depends crucially on the high-$x$ tail of $p(x)$, and the dissimilarity of the searched-for and background signals helps ensure a very steep fall-off for $p(x)$.

We emphasize, however, that LISA data analysis will also present confusion noise problems where there is no such separation of timescales. 
For instance, consider the search for relatively nearby EMRIS signals embedded in the background noise from all the unresolvably distant EMRIs. 
In that case the parent distribution $p(x_i)$ would surely
have a substantial tail, due to cases where $\delta \dot f_i$ is coincidentally small.
%, and the results of large-deviations would very likely not apply.
% In this case, the searched-for and background signals clearly have the same character, and one would expect $p(x)$ to have a quite substantial tail.  
Additionally, that detection problem raises issues of principle that we were not forced to confront 
in the model problems considered in this paper, and which we do not yet see how to resolve. 
For example, consider a case where some detection template $\tilde A$ has overlap of $5$, $4$, $3$ and $2$ with background signals A, B, C, and D, respectively.  Then the total SNR is 14 (assuming the sum of all other overlaps can be neglected),  which naively might lead one to claim a detection.  Should one consider that claim as a false alarm? What if most (but not all) of the parameters characterizing $\tilde A$ are fairly close to those of A?  Presumably experience with analyzing large sets of simulated data, as in the current Mock LISA Data Challenges, will alert us if such issues arise very often in practice. However if such issues arise only rarely, then our experience with this project suggests that
a sound theoretical understanding of the tails of the distribution could be crucial, 
since even with powerful computer clusters it could be difficult to sample the tails adequately with 
simulations.

\begin{acknowledgements}
\'{E}.R. was supported by NASA ATP grant NNG04GK98G awarded to E. Sterl Phinney.  C.C.'s work was carried out at the Jet Propulsion Laboratory, California Institute of Technology, under contract to the National Aeronautics and Space Administration.
\end{acknowledgements}

\appendix

\section{Heuristic introduction to large-deviations theory and correspondance with Edgeworth expansion}\label{LDapp}

Here we give a heuristic derivation of Chernoff's formula in large-deviations theory and discuss its
relation to the Edgeworth expansion. As before, let the sample mean be

\be
Z = \frac{1}{N}\sum_{i=1}^N x_i.
\ee
Now consider the modified cumulant generating functional $\Lambda(\tilde{\beta})$ for the sample mean PDF $P_N(Z)$. It is given by

\bea
\Lambda(\tilde{\beta}) &=& \ln \int e^{\tilde{\beta} Z} P_N(Z) dZ \nn \label{laplace1}\\
%&=& \log E\left[\exp\left(\tilde{\beta}\frac{1}{N}\sum_{i=1}^N x_i\right)\right] \nn \\
%&=& \log \prod_{i=1}^N E\left[\exp\left(\frac{\tilde{\beta}}{N}x_i\right)\right] \nn \\
&=& N \ln \int e^{\tilde{\beta}x/N} p(x) dx  \nn \\
&=& N \lambda(\beta) \, ,
\eea
where $\beta = \tilde{\beta} / N$ and where $\lambda$ is the modified cumulant generating functional of the parent distribution. Now $e^{\tilde\beta Z}$ is a rapidly increasing function of $Z$, while $P_N(Z)$ is rapidly decreasing.
Therefore one expects the integrand to be sharply peaked, and the integral to be some constant $C$ of order one times that the value of the integrand at that maximum.  [Of course, this is just Laplace's method of estimating the integral (\ref{laplace1}).]   Define $S(Z) \equiv - \ln P_N(Z)$.  Then we have just argued that $\Lambda(\tilde\beta)$ is well approximated by 
\be
\Lambda(\tilde\beta) = \max_Z\{\tilde\beta Z - S(Z)\} \, .
\ee
That is, $\Lambda(\tilde\beta)$ is the Legrendre transform of $S(Z)$, which we can invert to obtain
\bea
S(Z) &=& \max_{\tilde\beta}\{Z\tilde\beta - \Lambda(\tilde\beta)\} \\
&=&N\, \max_{\beta}\{Z\beta - \lambda(\beta)\} \, .
\eea
Alternatively we may write 
\be
P_N(Z) = C\,e^{-N I(Z)}
\ee
where
\be
I(Z) = \max_{\beta}[\beta Z - \lambda(\beta)].
\ee
and $C$ is a normalization constant determined {\it a posteriori}.   Clearly $C$ is approximately
given by $C \approx [N I^{\prime\prime}(0)/(2\pi)]^{1/2}]^{1/2}$.
This concludes our heuristic derivation of Chernoff's formula. 

As a pedagogical example, consider the random variable $X$ defined as

\be\label{RV}
X = \sum_{i=1}^N x_i,
\ee
where $x_i$ is a random variable equalling $+1$ or $-1$ with equal probability. The exact probability distribution for $X$ is a binomial, i.e.

\bea
P_N(X) &=& \frac{N!}{\left(\frac{N+X}{2}\right)!\left(\frac{N-X}{2}\right)!}\left(\frac{1}{2}\right)^{N+1} \nn \\
&=&  \frac{N!}{\left[\frac{N}{2}\left(1 + \frac{Y}{\sqrt{N}}\right)\right]!\left[\frac{N}{2}\left(1 - \frac{Y}{\sqrt{N}}\right)\right]!}\left(\frac{1}{2}\right)^{N+1}  ,\label{PofX}
\eea
where $Y = X/\sqrt{N}$ measures how many standard deviations away from the mean the variable $X$ lies. As $N$ tends to infinity, we make use of the following refined version of Stirling's formula \cite{stirling} to approximate (\ref{PofX}) as

\be\label{eq:stirling}
n! = \sqrt{2\pi}\, n^{n+1/2} e^{-n}\exp\left[\frac{1}{12n} - \frac{\theta_n}{360n^3}\right],
\ee
where the $\theta_n$ are all bounded between $0$  and $1$. By substituting (\ref{eq:stirling}) into (\ref{PofX}) and using $Z = Y/\sqrt{N}$, we obtain

\bea
P_N(Z) &=& \frac{1}{\sqrt{2\pi N}}\left(1 + Z\right)^{-N(1 + Z)/2 - 1/2}\left(1 - Z\right)^{-N(1 - Z)/2 - 1/2} \nn \\
&& \times \exp\left[\frac{1}{12N}\left(1 - \frac{2}{(1 + Z)} - \frac{2}{(1-Z)}\right) - \frac{1}{360N^3}\left({\theta_N} - \frac{8\theta_{N(1+Z)/2}}{(1+Z)^3} - \frac{8\theta_{N(1-Z)/2}}{(1-Z)^3}\right)\right] \nn \\\label{eq:PX1}
\eea

By further expanding in terms of $Z \ll 1$ and keeping the leading order corrections in $1/N$, we obtain
\bes
\bea
P_N(X) &=& \frac{1}{\sqrt{2\pi N}} e^{-\frac{X^2}{2N}}\exp\left\{\frac{1}{N}\left(-\frac{1}{4} + \frac{X^2}{2N} - \frac{X^4}{12N^2}\right)\times\left[1 + O(N^{-1})\right]\right\} \, , \label{eq:PXexpand1} \\
P_N(Y) &=& \frac{1}{\sqrt{2\pi}} e^{\frac{-Y^2}{2}}\exp\left\{\frac{1}{N}\left(-\frac{1}{4} + \frac{1}{2}Y^2 - \frac{1}{12}Y^4\right)\times\left[1 + O(N^{-1})\right]\right\} \, .  \label{eq:PXexpand2} 
\eea
\ees
\indent
Let us now derive the large-deviations prediction for $P_N(Z)$. First the cumulant generating functional is given by

\bea
\lambda(\beta) &=& \ln E[e^{\beta x}] \nn \\ 
&=& \ln(\cosh \beta).
\eea
Maximizing $I(Z)$ then yields

\be
Z = \frac{d\lambda(\beta)}{d\beta} = \tanh \beta.
\ee
The rate function is therefore given by

\bea
I(Z) &=& Z \,{\rm arctanh}\,Z - \ln [\cosh({\rm arctanh}\,Z)] \nn \\ 
&=&  \frac{1}{2}(1+Z)\ln(1+Z) + \frac{1}{2}(1-Z)\ln(1-Z).
\eea
This yields
\bea
P_N(Z) &=& C (1+Z)^{-N(1+Z)/2}(1-Z)^{-N(1-Z)/2}
\eea
for some normalization constant $C$.
Comparing this with (\ref{eq:PX1}), we see that it matches exactly the first line of (\ref{eq:PX1}), neglecting the small $-1/2$ term in each exponent. Large-deviations theory however does not capture the higher-order correction terms provided by that small $-1/2$ term and the entire second line of (\ref{eq:PX1}), which one needs to obtain expansion (\ref{eq:PXexpand1}).  
%[[Curt might also add section on stat mech.--or at least 2 sentences and a reference.]]
\indent
How does  large-deviations theory  "fit in" with the Central Limit Theorem and Edgeworth expansion?
%Presumably $N I(z)$ is the first term in an expansion in $1/N$.  
For simplicity let us assume that 
$p(x)$ is an even function (i.e., $p(-x) = p(x)$); clearly $P_N(Z)$ is then also even.
Presumably the exponent $N I(Z)$ appearing in Chernoff's formula is simply the lowest-order
term in an expansion in $1/N$:
\be\label{higherorder}
P_N(Z) =  \big[N \frac{d^2I}{dZ^2}(0)/(2\pi)\big]^{1/2}e^{-N \big(I(Z) + N^{-1} J(Z) + N^{-2} K(Z) + \cdots \big)}\, . %\ \ {\rm  as}\ N \rightarrow \infty.
\ee
Now expand each of $I(Z)$, $J(Z)$, and $ K(Z)$ as a power series  in $Z$:
% so that these expansions will contain only even powers
%of $Z$:
\bes
\bea
I(Z) &=& \big( i_2 Z^2 + i_4 Z^4 + \cdots \big), \\
J(Z) &=& \big(j_0 + j_2 Z^2 + j_4 Z^4 + \cdots \big), \\
K(Z) &=& \big(k_0 + k_2 Z^2 + k_4 Z^4 + \cdots \big), 
\eea
\ees
where the constants $j_0,k_0,\cdots$ are required for properly normalizing $P_N(Z)$ at each order in $N$. (There is no term $i_0$ in the expansion of $I(Z)$  because  the prefactor $\big[N \frac{d^2I}{dZ^2}(0)/(2\pi)\big]^{1/2}$ in Eq.~(\ref{higherorder}) ensures that $P_N(Z)$ is already normalized at lowest order. Of course, $i_2 = \frac{1}{2} \frac{d^2 I}{dZ^2}(0)$.)
%Here we have set $C = e^{-Ni_0}$, i.e. we have absorbed the leading order normalization constant into the definition of $I(Z)$. 
%The actual rate function given by Chernoff's formula satisfies by construction $I(0) = 0$.] 
Using $Y^2 \equiv  N Z^2$, we can then re-write Eq.~(\ref{higherorder}) as 
\bea\label{higherorderexp}
P_N(Y) &=& e^{- j_0}e^{-i_2Y^2}\exp\left[-\frac{1}{N}(k_0 + j_2Y^2 + i_4Y^4) + O(N^{-2})\right] \nn \\
&=& e^{- j_0}e^{-i_2Y^2}\left[1 -\frac{1}{N}(k_0 + j_2Y^2 + i_4Y^4) + O(N^{-2})\right]  \, . %\ \ {\rm as} \ N \rightarrow \infty. 
%P_N(Y) \rightarrow  e^{-Y^2 \big[i_0 + N^{-1}( j_0 + i_2 Y^2)  + N^{-2}(k_0 + j_2 Y^2 + i_4 Y^4) + 
%\cdots \big]}\ \  {\rm  as} \ Y \rightarrow \infty 
\eea
In this form the correspondence with the Edgeworth expansion becomes clear. The normalization of $P_N(Z)$ is fixed by $j_0$ and $k_0$ to that order, and the $i_2$ term represents the Central Limit Theorem
result, with $i_2 = (2\sigma_x^2)^{-1}$, and the $1/N$ terms are the leading-order corrections predicted by the Edgeworth series. For each term in the Edgeworth expansion, there is a piece that dominates at large $Y$.  Of course, this is  the term that contains the highest power of $Y$, e.g., the term 
$i_4 Y^4$ in Eq.~(\ref{higherorderexp}).  Large-deviations theory can be thought of as a clever way of summing up all these terms to determine
 the dominant large-$Y$ behavior. 
 
 Let us now show how this connection works for the binomial distribution. In that case, the first two cumulants are easily shown to be

\be
\kappa_2 = 1, \ \ \ \ \  \kappa_4 = -2.
\ee
%\ees
Thus its Edgeworth expansion is given by [cf. Eq.(\ref{Edgeworthexplicit})]

\be\label{eq:binEdg}
P_N(Y) = \frac{1}{\sqrt{2\pi}}e^{-Y^2/2}\left[1 - \frac{1}{12N}(Y^4 - 6Y^2 + 3) \right].
\ee
If one expands the $N$-dependent exponential in (\ref{eq:PXexpand2}) to leading order, one obtains precisely (\ref{eq:binEdg}). Now from (\ref{eq:PX1}), we may identify the functions $I(Z), J(Z)$ and $K(Z)$ appearing in (\ref{higherorder}) as

\bes
\bea
I(Z) &=&  %-\frac{1}{2}\log(2\pi N) +
 \frac{1}{2}(1+Z)\log(1+Z) + \frac{1}{2}(1-Z)\log(1-Z), \\
J(Z) &=& \frac{1}{2}\log(1+Z) + \frac{1}{2}\log(1-Z), \\
K(Z) &=& -\frac{1}{12}\left(1-\frac{4}{1-Z^2}\right).  
\eea
\ees
By expanding each of these functions around $Z=0$, we obtain $i_2 = 1/2$, $i_4 = 1/12$, 
$j_0 = 0$, $j_2 = -1/2$, and $k_0 = 1/4$.  Substituting these values into  (\ref{higherorderexp}), we recover precisely Eq.(\ref{eq:binEdg}). 
%This example illustrates the connection spelled out above between large-deviations theory and Edgeworth expansion.

\end{document}